\documentclass[
  aps,prx,twocolumn,nofootinbib,
  longbibliography,amsfonts,amssymb,amsmath,
  superscriptaddress, showkeys
]{revtex4-2}

\usepackage{amsthm}
\usepackage[USenglish]{babel}
\usepackage{bbm}
\usepackage{bm}
\usepackage{enumitem}
\usepackage{graphicx}
\usepackage[pdftex,colorlinks=true,allcolors=blue1]{hyperref}
\usepackage{lmodern}
\usepackage{mathtools}
\usepackage{siunitx}
\DeclareSIUnit{\molar}{M}
\usepackage{stackengine}
\usepackage[caption=false]{subfig}
\usepackage{tikz}
\usetikzlibrary{arrows.meta, positioning, shapes.multipart, calc, decorations.pathreplacing, fit}
\usepackage[dvipsnames]{xcolor}

\usepackage{xr}
\newcommand*{\addFileDependency}[1]{
  \typeout{(#1)}
  \IfFileExists{#1}{}{\typeout{No file #1.}}
}
\newcommand*{\myexternaldocument}[1]{
    \externaldocument{#1}
    \addFileDependency{#1.tex}
    \addFileDependency{#1.aux}
}
\myexternaldocument{sm}

\definecolor{blue1}{HTML}{1065AB} % blue1 % R 016 G 101 B 171
\definecolor{red2}{HTML}{B31529} % red2   % R 179 G 021 B 041
\definecolor{darkred}{rgb}{0.55, 0.0, 0.0}
\definecolor{darkblue}{rgb}{0.0, 0.0, 0.55}
\definecolor{darkgreen}{rgb}{0.0, 0.55, 0.0}

\newcommand{\panel}[3][]{%
  \begin{tikzpicture}[baseline=(img.north)]
    \node[anchor=south west, inner sep=0] (img)
      {\includegraphics[#1]{#3}};
    \node[anchor=base west, xshift=-1pt, yshift=0pt] % increased due to figure 7a and b
      at (img.north west) {\large \sf \textbf{#2}};
    \end{tikzpicture}%
}

\usepackage[normalem]{ulem} % To cross out text with \sout

\begin{document}

\title{Gradient estimators for parameter inference in discrete stochastic kinetic models}

\author{Ludwig Burger}
\email{ludwig.burger@tum.de}
\affiliation{Physics of Complex Biosystems, School of Natural Sciences, James-Franck-Stra\ss e 1, 85748 Garching, Germany}

\author{Annalena Kofler}
\affiliation{Max Planck Institute for Intelligent Systems, Max-Planck-Ring 4, 72076 T\"ubingen, Germany}
\affiliation{Max Planck Institute for Gravitational Physics (Albert Einstein Institute), Am M\"uhlenberg 1, 14476 Potsdam, Germany}

\author{Lukas Heinrich}
\affiliation{Data Science in Physics, School of Natural Sciences, James-Franck-Stra\ss e 1, 85748 Garching, Germany}
\affiliation{Munich Center for Machine Learning (MCML), Munich, Germany}

\author{Ulrich Gerland}
\email{gerland@tum.de}
\affiliation{Physics of Complex Biosystems, School of Natural Sciences, James-Franck-Stra\ss e 1, 85748 Garching, Germany}

\begin{abstract}
Stochastic kinetic models are ubiquitous in physics, yet inferring their parameters from experimental data remains challenging. 
For deterministic models, parameter inference often relies on gradients, which can be obtained efficiently through automatic differentiation (AD). 
However, AD cannot be applied directly to the Gillespie stochastic simulation algorithms~(SSA), since sampling from a discrete set of reactions introduces non-differentiable operations. 
In this work, we adopt three gradient estimators from machine learning for the Gillespie~SSA: the Gumbel-Softmax Straight-Through~(GS-ST) estimator, the Score Function estimator, and the Alternative Path estimator.
We use the estimators to evaluate gradients of steady-state and time-dependent observables, and compare their performance in representative biophysical systems with relaxation dynamics (bimolecular association) and oscillatory dynamics (repressilator).
We find that the GS-ST estimator generally yields well-behaved gradient estimates, but exhibits diverging variance in challenging parameter regimes, which can cause parameter inference to fail.
In these cases, other estimators provide more robust, lower variance gradients.
Our results demonstrate that gradient-based parameter inference can be effectively combined with the Gillespie~SSA, with different estimators offering complementary advantages.
\end{abstract}

\maketitle 

%%%%%%%%%%%%%%%%%%%%%%%%%%%%%%%%%%%%%%%%%%%%%%%%%%%%%%%%%%%%%%%%%%%%%%%%%%%%%%%%
%%%%%%%%%%%%%%%%%%%%%%%%%%%%%%%%% introduction %%%%%%%%%%%%%%%%%%%%%%%%%%%%%%%%%
%%%%%%%%%%%%%%%%%%%%%%%%%%%%%%%%%%%%%%%%%%%%%%%%%%%%%%%%%%%%%%%%%%%%%%%%%%%%%%%%
\section{Introduction}

Inferring unknown model parameters from experimental data is a fundamental challenge across the physical sciences~\citep{brehmer:2019xox, khan2021bayesian, carter2024benchmark, kofler2025flexible, gavrikov2026neutrinosbi}. 
Parameter inference becomes particularly demanding when the dimension of the parameter space is high, or when relevant aspects of the dynamics can only be inferred from incomplete observations. 
These challenges are typical of complex biochemical systems, whose behavior is governed by many coupled biochemical reactions, while only a subset of molecular species can be monitored experimentally. 
For chemical reaction networks described by deterministic rate equations, a wide range of methods has been developed both for obtaining best-fit parameters and for quantifying their uncertainty~\citep{mitra2020parameter}. 
In particular, gradient-based inference methods have become standard tools, since derivatives of the objective function with respect to model parameters can be computed by finite difference estimates, forward or adjoint sensitivity analysis~\citep{froehlich2017scalable,stapor2018optimization}, or automatic differentiation~\citep{frank2022automatic}.

However, in many biochemical systems, low copy numbers and the resulting intrinsic fluctuations limit the applicability of deterministic models and necessitate a stochastic description~\citep{arkin1998stochastic,schnoerr2017approximationandinference}. 
The corresponding stochastic kinetic models are commonly formalized as continuous-time Markov chains, and are typically simulated using the Gillespie stochastic simulation algorithm~(SSA)~\citep{gillespie1977exact}. 
Parameter inference for such stochastic models is substantially more challenging~\citep{komorowski2011sensitivity,ruess2017sensitivity}. 
Among the most widely used strategies are moment-based inference, often implemented using moment-closure approximations~\citep{zechner2012momentbased, schnoerr2015comparison, lueck2016generalized,froehlich2016inference}, and likelihood-based inference, which relies on tractable approximations of the likelihood~\citep{komorowski2009bayesian} or on simulation-based likelihood estimation~\citep{golightly2011bayesian}.
Likelihood-free inference, such as approximate Bayesian computation~\citep{toni2009abc,warne2019simulation}, provides an alternative whenever the likelihood is intractable. 
Unlike parameter inference in deterministic models, gradient-based approaches are less straightforward for stochastic kinetic models simulated with the Gillespie~SSA, because trajectories are generated by sampling discrete reaction events and waiting times from parameter-dependent probability distributions.

The challenge of estimating gradients in stochastic simulations with discrete randomness also arises in several machine learning settings, for example in reinforcement learning and generative modeling~\citep{mohamed2020montecarlo}. 
This has led to the development of a range of gradient estimators, including the Gumbel-Softmax Straight-Through~(GS-ST) estimator~\citep{jang2017categorical,maddison2017discrete}, the Score Function~(SF) estimator~\citep{williams1992simple}, and the Alternative Path~(AP) estimator~\citep{arya2022automatic}. 
These gradient estimators have been used for parameter inference or inverse design tasks in several branches of physics, including particle physics or statistical physics~\citep{kagan2023branches, heinrich2026differentiablequantum, arya2023diff}.
\looseness=-1

In this paper, we adapt three gradient estimation techniques (GS-ST, SF, and AP) for parameter inference in the Gillespie SSA. 
First, we assess their performance in a time-independent steady-state setting, focusing on variance as a function of system parameters, trajectory length, and estimator-specific parameters.
Second, we generalize the gradient estimators to time-dependent settings and examine how including gradient contributions from the stochastic waiting times affects estimator performance.
Third, we apply the gradient estimators to parameter inference in a stochastic repressilator system exhibiting oscillatory dynamics~\citep{elowitz2000syntheticoscillatory, loinger2007stochastic, potvin2016synchronous}.

While concurrent studies also demonstrate that GS-ST~\citep{mottes2026gradient,vilar2026exact} and other relaxation-based gradient estimators~\citep{rijal2025diffablegillespie} can in principle be used for parameter inference, we identify parameter regimes and estimator settings in which inference via GS-ST fails due to diverging variance and strong bias. 
We show that the unbiased SF~estimator facilitates robust inference, even in these challenging cases.
Overall, our results indicate that gradient-based optimization is a promising avenue for parameter inference in stochastic models, with different estimators exhibiting complementary strengths and limitations.

%%%%%%%%%%%%%%%%%%%%%%%%%%%%%%%%%%%%%%%%%%%%%%%%%%%%%%%%%%%%%%%%%%%%%%%%%%%%%%%%
%%%%%%%%%%%%%%%%%%%%%%%%%%%% steady-state estimators %%%%%%%%%%%%%%%%%%%%%%%%%%%
%%%%%%%%%%%%%%%%%%%%%%%%%%%%%%%%%%%%%%%%%%%%%%%%%%%%%%%%%%%%%%%%%%%%%%%%%%%%%%%%
\section{Gradient Estimation in Gillespie~SSA for Steady-State Observables}
In the Gillespie SSA, the time evolution of a chemical reaction network is represented as a sequence of discrete reaction events separated by random waiting times~\citep{gillespie1977exact}.
At each step~$s$, the waiting time $\Delta t_s$ until the next reaction and the chemical reaction that occurs after the waiting time are sampled. 
The sampled reaction channel is denoted by its index~$I_s$ in the set of allowed reactions and uniquely determines the change in molecule count~$\Delta N_s$.
The time $t_{s-1}$ and molecule count $N_{s-1}$ are updated according to the sampled waiting time and reaction channel index,
\begin{align*}
    (N_{s-1}, t_{s-1}) \quad \xrightarrow[\Delta t_s]{I_s\,\Rightarrow\,\Delta N_s} \quad (N_s, t_s) \, .
\end{align*}
The probability~$p(I_s \mid N_{s-1})$ of selecting reaction channel~$I_s$ is proportional to its propensity, which depends on the current molecule count~$N_{s-1}$. 
Here, $N_{s-1}$ is a vector with dimensionality equal to the total number of molecular species in the system.
Since the molecule count~$N_{s-1}$ at step~$s-1$ is fully determined by the reaction history, i.e., the sequence of previously sampled reaction channel indices $I_{\leq s-1} = \{I_1, \dots, I_{s-1} \}$, we may equivalently write $p(I_s \mid N_{s-1}) = p(I_s \mid I_{\le s-1})$.
Based on the sampled reaction index~$I_s$ at step~$s$, the molecule count is updated according to $N_{s} = N_{s-1} + \Delta N_s$, with vector~$\Delta N_s$ denoting the change in molecule counts induced by the reaction event. 
Explicitly, $\Delta N_s = \mathcal{S} X_s$, where $X_s$ is a vector that is non-zero only at the sampled reaction index~$I_s$ and zero otherwise, and~$\mathcal{S}$ denotes the stoichiometric matrix encoding the changes in molecule counts associated with each reaction channel~\citep{rao2016nonequilibrium}.
The waiting time~$\Delta t_s$ until the next reaction is drawn from an exponential distribution $p(\Delta t_s \mid N_{s-1}) = p(\Delta t_s \mid I_{\le s-1})$ with a characteristic timescale set by the total propensity~$a^\mathrm{tot}_{s-1}$, which in turn depends on the current molecule count. 
The time is then incremented by the sampled waiting time, $t_s = t_{s-1} + \Delta t_s$. 
This procedure is repeated until the accumulated simulation time~$t_s$ reaches the prescribed final time~$t_\mathrm{final}$.

The full Gillespie SSA generates samples from the time-dependent distribution
$p(N,t)$, allowing us to evaluate the (time-dependent) expectation value $\langle f(N) \rangle_t$ of an observable $f(N)$.
However, to compute the steady-state expectation value, we require averages with
respect to the steady-state distribution $p_{\rm s.s.}(N)$. 
One approach to obtain such samples is to run the Gillespie SSA to a final time $t_{\rm final}$ much larger than the equilibration time, at which point $N$ is approximately distributed according to $p_{\rm s.s.}(N)$.
Alternatively, we can sample the molecule count $N_s$ after $s$ Gillespie steps and reweight each sample by the mean lifetime of the corresponding
state $(a^{\rm tot}(N_s))^{-1}$~\citep{anderson2017stochastic}. 
In this case, $s$ must be chosen sufficiently large that the distribution of $N_s$ has reached stationarity.
The steady-state expectation value of an observable $f(N)$ is then
\begin{align}
    \langle f(N)\rangle_{\rm s.s.}
    =
    \frac{
        \left\langle f(N_s)\, [a^{\rm tot}(N_s)]^{-1} \right\rangle_s
    }{
        \left\langle [a^{\rm tot}(N_s)]^{-1} \right\rangle_s
    } \, ,
    \label{eq:steady_state_average}
\end{align}
where $\langle \cdot \rangle_{\rm s.s.}$ denotes the average over the full
steady-state distribution, and $\langle \cdot \rangle_s$ the average over
the stationary distribution of $N_s$ reached after $s$ Gillespie steps.

While the Gillespie SSA provides an exact and efficient way to generate samples for a given set of parameters~$\theta$ (e.g. rate constants of chemical reactions), the corresponding inverse problem of inferring parameters from samples is considerably less straightforward. 
Determining the parameter values that best reproduce the behavior observed in a set of samples requires information about the sensitivity of the simulation output to the input parameters, for instance through gradients such as $\nabla_\theta \langle f(N)\rangle_t$ or $\nabla_\theta \langle f(N) \rangle_{\rm s.s.}$. 
However, evaluating such gradients is challenging for two reasons.
First, the expectation values depend on the parameters through the probability distributions $p(N, t \mid \theta)$ or $p_{\rm s.s.}(N \mid \theta)$.
Second, the dynamics are governed by discrete reaction events in a discrete state space, which makes gradients difficult to define.

We consider three gradient estimators for use with the Gillespie SSA:
(i)~the GS-ST,
(ii)~the SF, and
(iii)~the AP estimator.
All three methods estimate gradients of expectation values with respect to the parameters~$\theta$ by constructing a gradient estimator $\mathcal{G}$ such that
\begin{align*}
    \nabla_\theta \langle f(N) \rangle = \langle \mathcal{G}[f(N)] \rangle \, .
\end{align*}
The precise form of~$\mathcal{G}$ differs between the estimators and is detailed in the following. 
For all estimators, we distinguish two separate scenarios, namely gradient estimation of
(i)~steady-state observables, $\nabla_\theta \langle f(N) \rangle_{\rm s.s.}$, 
and (ii)~time-dependent observables, $\nabla_\theta \langle f(N) \rangle_{t}$. 
In the former case, the observable is evaluated via Eq.~(\ref{eq:steady_state_average}), such that only the molecule count after $s$ steps, $N_s$, enters the estimator, whereas the waiting times between reactions do not.
Consequently, for steady-state observables, the sampling of waiting times and their gradient contribution can be neglected.
For time-dependent observables, by contrast, the gradient contribution of the waiting times must be taken into account explicitly.

\subsection{Definition of Gradient Estimators}

In the following, we first introduce each estimator for a single categorical random variable $I$ distributed according to~$p(I \mid \theta)$.
We then extend the discussion to sequences of categorical random variables.
For such sequences, we assume that the distribution of $I_s$ at step~$s$, written as $p(I_s \mid I_{\le s-1}; \theta)$, depends on the full history $I_{\le s-1} = \{I_1, \ldots, I_{s-1}\}$, mirroring the structure of trajectories generated by the Gillespie~SSA for a fixed number of steps. 
The generalizations required for time-dependent observables, where the number of steps is stochastic, are discussed in a dedicated section later in the manuscript. \\

\paragraph*{Gumbel-Softmax Straight-Through Estimator.}
The Gumbel-Softmax Straight-Through (GS-ST) estimator~\citep{jang2017categorical,maddison2017discrete} is based on the Gumbel--max trick, which provides a reparameterized representation of sampling from a categorical distribution~\citep{gumbel1954statistical}. 
A sample of a categorical random variable is obtained by drawing independent random variables $g_i$ from a standard Gumbel distribution with probability density function $\exp\left[-g_i - {\rm e}^{-g_i} \right]$, and evaluating
{\fontsize{9.5}{11}\selectfont 
\begin{align}
    X_i = 
    \begin{cases}
        1 & \text{if} \, \, \log p(I = i) + g_i - \log p(I = j) - g_j > 0 \quad \forall j \, , \\
        0 & \text{else}.
    \end{cases}
    \label{eq:gsst__fwd}
\end{align}
}\noindent
Thus, $X$ denotes a vector that is equal to one only for the sampled category~$I = i$ and zero otherwise, a representation commonly referred to as one-hot encoding~\citep{harris2012digital, zheng2018feature}. 
Since the components of~$X$ have discrete values, gradients with respect to the parameters $\theta$ cannot be evaluated directly. 
To obtain a differentiable expression, $X_i$ is approximated as a `softmax' function 
\begin{align*}
    X_i^\mathrm{relaxed} 
    &= \frac{\exp\!\left[(\log p(I=i) + g_i)/\tau\right]}
         {\sum_j \exp\!\left[(\log p(I=j) + g_j)/\tau\right]} \, ,
\end{align*}
which, from a statistical physics perspective, can be regarded as a canonical probability distribution with temperature $\tau$ (the denominator then corresponds to the partition function).
$X_i^{\rm relaxed}$~approaches one when the noisy log-likelihood $\log p(I=i)+g_i$ \looseness=-1 is larger than all other noisy log-likelihood values, and zero when it is sufficiently smaller, with the sharpness of the crossover controlled by the temperature~$\tau$.
Using $X_i^{\rm relaxed}$, the GS-ST gradient estimator for the $i$th component is defined as 
\begin{align}
    \mathcal{G}_{\rm GSST}[f(I)]
    &= \nabla_\theta f(X_i^\mathrm{relaxed}) \, .
    \label{eq:gsst__bwd}
\end{align}
Since the stochasticity enters only via the parameter-independent Gumbel noise $g_i$, $\mathcal{G}_{\rm GSST}$ can be evaluated using standard automatic differentiation. 
This construction is commonly referred to as the reparameterization trick~\citep{rezende2014stochastic,kingma2014autoencoding}. 
The desired gradient ${\rm d} \langle f(I)\rangle / {\rm d} k$ is obtained by averaging $\mathcal{G}_{\rm GSST}$ over independent samples of the Gumbel noise. 
Importantly, the continuous relaxation $X_i^{\rm relaxed}$ is only used for gradient evaluation (backward pass), while the simulation output (forward pass) always uses the exact sample $X_i$ to retain an unaltered exact trajectory. 
The use of a continuous relaxation generally introduces a bias in the estimated gradient.
Increasing the temperature $\tau$ increases this bias but reduces the variance of the gradient estimate, giving rise to a bias-variance trade-off~\citep{jang2017categorical,maddison2017discrete}. 

In the Gillespie SSA, the time evolution of a chemical reaction network is described as a sequence of chemical reactions.
For the moment, we restrict attention to gradients of time-independent observables in steady state, for which the stochastic timing of reactions need not be considered (the general case is discussed in the section on time-dependent observables).
To sample the reaction channel, we use Eq.~(\ref{eq:gsst__fwd}) with probabilities determined by the reaction propensities, $p(I=i)=a^{(i)}_{s-1}/a_{s-1}^{\rm tot}$. 
This yields the vector $X_s$ indicating the sampled reaction channel, while the corresponding gradient $\nabla_\theta X_s$ is obtained via Eq.~(\ref{eq:gsst__bwd}). 
The resulting change in molecule count is then given by $\Delta N_s = \mathcal{S} X_s$, with gradient $\nabla_\theta \Delta N_s = \mathcal{S}\,\nabla_\theta X_s$, where $\mathcal{S}$ denotes the stoichiometric matrix. 
The molecule count after $s$ Gillespie steps is given by $N_s = N_0 + \sum_{r \leq s} \Delta N_r$. 
Since the gradients $\nabla_\theta \Delta N_r$ are known for all $r \leq s$ (and $N_0$ is assumed to be parameter-independent), $\nabla_\theta N_s$ can be obtained directly. \\

\paragraph*{Score Function Estimator.}
While the GS-ST estimator is subject to a $\tau$-dependent bias, the SF estimator provides an unbiased alternative~\citep{williams1992simple, mohamed2020montecarlo}. 
For a categorical random variable~$I$, the gradient of the expectation value can be rewritten as 
\begin{align*}
    \nabla_\theta \, \langle f(I) \rangle
    &= \nabla_\theta \sum_i f(i)\, p(I=i) \notag \\
    &= \sum_i f(i)\, p(I = i)\,
      \nabla_\theta \log p(I = i) \\
      &= \langle f(I) \, \nabla_\theta \log p(I) \rangle \, ,
\end{align*}
suggesting the estimator
\begin{align*}
    \mathcal{G}_{\rm SF}[f(I)]
    = \bigl(f(I)-b\bigr)\,
      \nabla_\theta \log p(I) \, ,
\end{align*}
where $\nabla_\theta \log p(I)$ is the score function, and $b$ denotes a sample-independent baseline. 
Subtracting the baseline reduces the variance without biasing the estimator; a common choice is the mean baseline, $b=\langle f(I)\rangle$~\citep{schulman2016gradient}.
For a sequence of random variables~$I_{\leq s} = \{ I_1, \dots, I_s \}$ distributed according to $p(I_{\leq s})$, this becomes 
{\fontsize{9.5}{11} \selectfont 
\begin{align}
    \mathcal{G}_{\rm SF}[f(I_{\leq s})]
    =
    \biggl(f(I_{\leq s})-\langle f(I_{\leq s})\rangle\biggr) \,
    \nabla_\theta \log p(I_{\leq s}) \, .
    \label{eq:SF}
\end{align}}\noindent
where $f(I_{\leq s})$ denotes an observable computed based on the samples up to step $s$.
Since $p(I_{\leq s})$ can be factorized into conditional distributions, the score function can be expressed as a sum of the conditional scores,
\begin{align*}
    \nabla_\theta \log p(I_{\leq s}) = \sum_{r \le s}
    \nabla_\theta \log p(I_r \mid I_{<r}) \, .
\end{align*}

In the Gillespie SSA, the observable of interest $f(I_{\leq s})$ is typically a function of the molecule count, $N_s = N_0 + \sum_{r \leq s} \Delta N_r$, where each molecule-count update $\Delta N_r$ is uniquely determined by the sampled reaction channel $I_r$. 
Due to the unique correspondence between $I_s$ and $\Delta N_s$, the probability distributions $p(I_{\leq s})$ and $p(\Delta N_{\leq s})$ can be used interchangeably. Thus, the score function reads
\begin{align}
    \nabla_\theta \log p(\Delta N_{\leq s}) 
    &= \sum_{r \leq s} \nabla_\theta \log p(\Delta N_r \mid N_{r-1} ) \, , \label{eq:SF__molecule_score}
\end{align}
where we have used that the distribution of $\Delta N_r$ depends on the past only through the molecule count $N_{r-1}$, but is independent of the detailed trajectory $\Delta N_{< r}$ by which this state was reached.

The score function $\nabla_\theta \log p(\Delta N_{\leq s})$ based on the molecule-count updates is only sufficient to evaluate gradients of time-independent observables. 
For time-dependent gradients, the score function needs to include the molecule update as well as the waiting time, $\nabla_\theta \log p(\Delta N_{\leq s}, \Delta t_{\leq s})$, as discussed further below. \\

\paragraph*{Alternative Path Estimator.}
\label{sec:AP}
Like the SF estimator, the AP estimator~\citep{arya2022automatic} is an unbiased gradient estimator applicable to discrete random variables. 
Its conceptual idea is to couple a \emph{primal} sample drawn at parameter value~$\theta$ to an \emph{alternative} sample at a shifted parameter value $\theta + \epsilon$ by using the same source of randomness for both samples. 
The primal sample~$i$ is drawn from the categorical distribution $p(I;\theta)$.
To identify admissible alternative samples~$j$, it is useful to recall that categorical sampling can be implemented via inverse transform sampling: a~uniform random variable $u\sim \operatorname{Uniform}(0,1)$ is mapped to a category through decision boundaries determined by the cumulative probabilities.
An infinitesimal change $\theta\mapsto\theta+\epsilon$ shifts these decision boundaries.  
As a result, the sampled category changes when a decision boundary adjacent to the primal category crosses the fixed value of~$u$.
In the limit $\epsilon\to0$, this restricts the set of admissible alternative samples to the two neighboring categories $j = i-1$ or $j = i+1$. The ordering of the categories is arbitrary, but must remain fixed along the trajectory.

Rather than explicitly resampling the alternative output, the AP~estimator enumerates the admissible alternative samples and assigns boundary-shift weights to them,
\begin{align*}
    w_{-}(i)
    &= \frac{1}{p(i)}\,
    \frac{\rm d}{{\rm d} \theta}
    \sum_{i'<i} p(i')\,
    \mathbbm{1}\bigg(
    \frac{{\rm d}}{{\rm d} \theta}
    \sum_{i'<i} p(i') > 0
    \bigg) \, , \\[1ex]
    w_{+}(i)
    &= -\,\frac{1}{p(i)}\,
    \frac{{\rm d}}{{\rm d} \theta}
    \sum_{i' \leq i} p(i')\,
    \mathbbm{1}\bigg(
    \frac{{\rm d}}{{\rm d} \theta}
    \sum_{i' \leq i} p(i') < 0
    \bigg) \, ,
\end{align*}
which quantify how strongly the lower or upper decision boundary adjacent to the primal category~$i$ shift under a change in the parameter~$\theta$. 
Indicator functions~$\mathbbm{1}(\cdot)$ ensure that only boundary shifts moving into the interval corresponding to the primal category contribute; otherwise, the respective alternative path is not accessible. 
If there are multiple parameters, then each parameter $\theta_k$ has its own boundary shift weights. 
The alternative sample~$j$ is drawn from $\{i-1,i+1\}$ with probabilities proportional to the corresponding weights,
\begin{align*}
    p(j=i\pm1 \mid i) = \frac{w_{\pm}(i)}{w_{-}(i)+w_{+}(i)} \, .
\end{align*}
Defining the total weight as $w(i) = w_{-}(i) + w_{+}(i)$, the AP estimator for the gradient takes the form
\begin{align}
    \mathcal{G}_{\rm AP}[f(I)] = (f(J)-f(I))\, w(I) \, .
    \label{eq:AP}
\end{align}

For sequences of random variables, the AP estimator is applied iteratively along the sampled sequence.
At step~$s$, the primal sample~$i_s$ is drawn from the conditional distribution $p(I_s \mid I_{\le s-1} = i_{\leq s-1})$ by inverse transform sampling using a uniform random variable~$u_s$. 
For the alternative path, two options arise at each step.
First, a new alternative path may be generated by applying the procedure outlined above.
This yields an alternative sample~$j_s \neq i_s$ with an associated boundary-shift weight $w_s$.
Second, an existing alternative path~$j_{\le s-1}$ with weight $w_{\leq s-1}$ generated at an earlier step may be continued.
In this case, the alternative sample~$j_s$ is obtained by sampling from the distribution $p(I_s \mid I_{\le s-1} = j_{\leq s-1})$ using the \emph{same} uniform random variable~$u_s$ as for the primal path.
Importantly, the distribution is conditioned on the existing alternative path $j_{\leq s-1}$ instead of the primal path $i_{\leq s-1}$.
The estimator selects between generating a new alternative path at step~$s$ and continuing an existing one by sampling according to their relative weights.
Specifically, the new alternative is chosen with probability $w_s / w_{\le s}$, while the existing alternative is continued with probability $w_{\le s-1} / w_{\le s}$, where $w_{\le s} = w_{\le s-1} + w_s$. 
Once selected, the alternative path is propagated forward in the same manner at subsequent steps. 

In the Gillespie SSA, the AP procedure described above is used to obtain the gradient contribution associated with the sampling of the reaction channels. The gradient contribution of the stochastic waiting times is discussed in the section on time-dependent observables. 

%%%%%%%%%%%%%%%%%%%%%%%%%%%%%%%%%%%%%%%%%%%%%%%%%%%%%%%%%%%%%%%%%%%%%%%%%%%%%%%%
%%%%%%%%%%%%%%%%%%%%%%%%%%%%% steady state %%%%%%%%%%%%%%%%%%%%%%%%%%%%%%
%%%%%%%%%%%%%%%%%%%%%%%%%%%%%%%%%%%%%%%%%%%%%%%%%%%%%%%%%%%%%%%%%%%%%%%%%%%%%%%%
\subsection{Example: Bimolecular Association}

\begin{figure*}
    \subfloat[]{\panel[width=0.32\linewidth]{a}{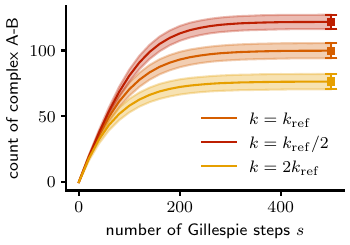}} \hfill
    \subfloat[]{\panel[width=0.32\linewidth]{b}{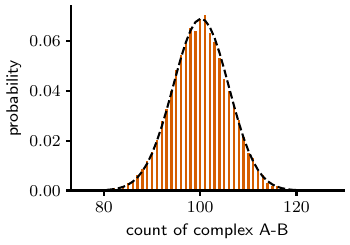}} \hfill
    \subfloat[]{\panel[width=0.32\linewidth]{c}{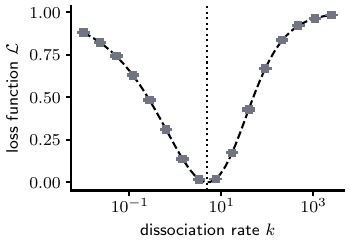}} \\[-5ex]
    \subfloat[]{\panel[width=0.32\linewidth]{d}{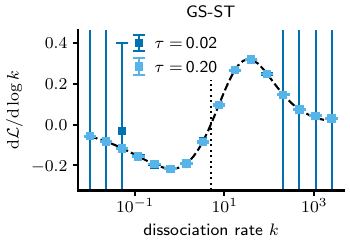}} \hfill
    \subfloat[]{\panel[width=0.32\linewidth]{e}{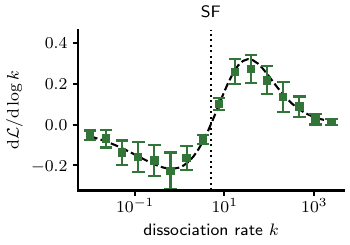}} \hfill
    \subfloat[]{\panel[width=0.32\linewidth]{f}{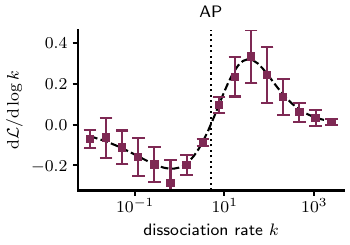}} \\[-3ex]
    \caption{Bimolecular association model.
    (a)~Starting from an initial configuration without bimolecular complexes, the system relaxes toward a steady state with a non-zero number of A--B complexes.
    The number of complexes in steady-state is governed by the dissociation rate $k$, the total numbers of molecules A and B, and the system volume
    (here $N_{\rm A}^{\rm tot}=N_{\rm B}^{\rm tot}=200$, $V=20$, and $k_{\rm ref}=5$).
    Solid lines and shaded areas show the mean and standard deviation of $10^4$ trajectories.
    (b)~The steady-state distribution of A--B complexes obtained after $s = 500$ Gillespie steps (orange bars) agrees well with the distribution derived from the exact solution of the Chemical Master Equation (dashed line, see Appendix~\ref{APsec:bima}).
    The histogram is obtained from $10^4$ samples, with each bin weighted by the lifetime of the corresponding state to obtain an unbiased steady-state distribution (Eq.~\ref{eq:steady_state_average}).
    (c)~The loss function, defined as the relative deviation between the average steady-state number of complexes at $k$ and at $k_{\rm ref}$, exhibits a clear minimum at~$k_{\rm ref}$.
    (d--f)~The gradient of the loss function, ${\rm d}\mathcal{L}/{\rm d}\log k$, evaluated using the GS-ST method~(d), the SF method~(e), and the AP method~(f), agrees well with the analytical reference (dashed lines).
    Data points in panels (c--f) show the mean and uncertainties obtained by averaging over 10~independent batches, each consisting of 100~stochastic trajectories.
    }
    \label{fig:bima_overview_step}
\end{figure*}

The estimators introduced so far provide trajectory-level gradient estimates for observables computed from the output of the Gillespie SSA.
To use these estimators for parameter inference, we must specify a loss function that quantifies the deviation between the simulation output and the (experimental) observations.
For steady-state observables, the most general loss function can be written as a functional of the probability distribution of the simulation output for a given choice of parameters, $p_{\rm s.s.}(N; \theta)$, and the distribution of the observed reference data $p_{\rm ref}(N)$,
\begin{align*}
\mathcal{L}(\theta) = \mathcal{L}[p_{\rm s.s.}(N; \theta), p_{\rm ref}(N)] \, .
\end{align*}
In practice, many choices for the loss function are possible, ranging from losses that match only a few relevant moments of the distribution to losses that compare the full distribution~\citep{mottes2026gradient}, such as the Kullback--Leibler divergence or the Earth Mover's distance~\citep{rubner2000earthmoversdistance}. 

As a simple example of a process that reaches a time-independent steady state, we consider the bimolecular association reaction, where two molecular species A~and~B reversibly form a bimolecular complex A-B, 
\begin{align*}
    {\rm A} + {\rm B} \;\xrightleftharpoons[k_{\rm off}]{k_{\rm on}}\; \text{A-B}\, .
\end{align*}
For this process, it is straightforward to compute the probability distribution over the entire state space, providing an exact reference against which the simulation output and its gradients can be compared (Appendix \ref{APsec:bima}). 
We assume that the total molecule numbers of~A and~B, $N_{\rm A}^{\rm tot}$ and $N_{\rm B}^{\rm tot}$, are conserved, and the system volume~$V$ is fixed; 
the number of bimolecular complexes~A-B after $s$ Gillespie steps is denoted by $N_s$.
For simplicity, we set the association rate to $k_{\rm on}=1$ and treat the dissociation rate~$k_{\rm off} = k$ as the only free parameter.
The system converges to a steady-state distribution (Fig.~\ref{fig:bima_overview_step}a--b),
and the dissociation rate~$k$ controls the steady-state number of complexes, with smaller values of~$k$ resulting in a higher number of A-B. 
The steady-state distribution is obtained by sampling $N_s$ via the Gillespie SSA, binning the outcomes, and weighting each bin in the histogram by the lifetime of the corresponding state (analogous to Eq.~\eqref{eq:steady_state_average}). 
The total number of Gillespie steps $s$ is chosen sufficiently large to ensure that the stationary distribution of $N_s$ is reached for all parameter values considered.

To compare the simulation output at parameter $k$ to the reference data (generated using the Gillespie SSA at $k = k_{\rm ref}$), we choose a loss function that measures the relative deviation of the average complex number from its reference value,
\begin{align}
    \mathcal{L}(k) = \left( \frac{\langle N(k) \rangle_{\rm s.s.} - \langle N(k_{\rm ref}) \rangle_{\rm s.s.}}{ \langle N(k_{\rm ref}) \rangle_{\rm s.s.}}\right)^2 \, .
    \label{eq:loss_bima_steady-state}
\end{align}
This loss function matches the first moments of the distribution, which is sufficient to infer the dissociation rate~$k$, if the remaining parameters (such as the total molecule numbers and system volume) are fixed. 
In principle, the loss function can be extended to include higher moments, for example to infer the system volume from the variance of the molecule number.

The loss~$\mathcal{L}(k)$ obtained from the Gillespie simulations agrees well with the loss computed from the exact solution of the Master equation (Fig.~\ref{fig:bima_overview_step}c, see Appendix \ref{APsec:bima}).
Using the three gradient estimators, we evaluate the gradient of the loss function, ${\rm d} \mathcal{L} / {\rm d} \log k$, and its uncertainty via a multi-sample estimate (Fig.~\ref{fig:bima_overview_step}d--f, see Supplementary Material Section~\ref{SMsec:sso__loss_gradient}).
Specifically, each gradient estimate is obtained by averaging gradients estimated from 100 independent stochastic trajectories; mean and standard deviation (data points and error bars in Fig.~\ref{fig:bima_overview_step}d--f) are then computed across 10 such estimates. 
We find that all three methods recover the correct gradient on average, but differ substantially in their uncertainty.
The GS-ST estimator yields a low-variance gradient estimate at high temperature (light blue data points in Fig.~\ref{fig:bima_overview_step}d), but its variance increases markedly as $\tau$ is reduced (dark blue data points), especially for small and high dissociation rates. The error bars significantly exceed the plotted $y$-range ($\sigma({\rm d} \mathcal{L}/{\rm d} \log k) \sim 10^{10}$). 
The SF~and AP~estimators show higher uncertainty than GS-ST at high temperature across the entire parameter range, but unlike GS-ST at low temperature, they do not exhibit diverging variance. Between SF and AP, SF yields the lower-variance gradient estimates.

\subsection{Variance Scaling of Gradient Estimators}
As shown in Fig.~\ref{fig:bima_overview_step}d--f, the uncertainties of the estimated gradients depend on the choice of estimator and the system parameters.
For example, the GS-ST estimator suffers from diverging variances at extreme dissociation rates and small temperature values.
To characterize this effect more quantitatively, we focus on the variances of the estimators for the gradient ${\rm d}\langle N_s\rangle/{\rm d}k$, which provide a proxy for the uncertainty of the loss gradient, since ${\rm d} \mathcal{L}/{\rm d} k$ is proportional to ${\rm d} \langle N_s \rangle/{\rm d}k$. 
For this analysis, we initialize each Gillespie simulation by sampling from the steady-state distribution, so that both $\langle N_s \rangle$ and its gradient ${\rm d} \langle N_s \rangle / {\rm d} k$ are independent of $s$. \\

\begin{figure}
    \vspace{-10pt}
    \subfloat[]{\panel[width=0.48\linewidth]{a}{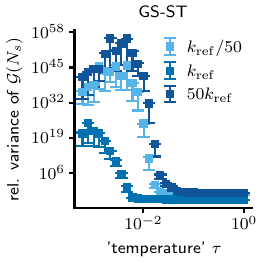}} \hfill
    \subfloat[]{\panel[width=0.48\linewidth]{b}{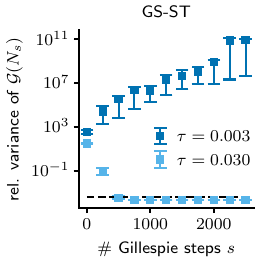}}
    \\[-5ex]
    \subfloat[]{\panel[width=0.48\linewidth]{c}{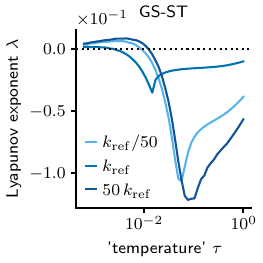}} \hfill
    \subfloat[]{\panel[width=0.48\linewidth]{d}{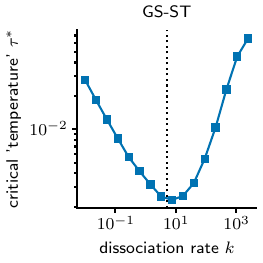}}
    \\[-3ex]
    \subfloat[]{\panel[width=0.48\linewidth]{e}{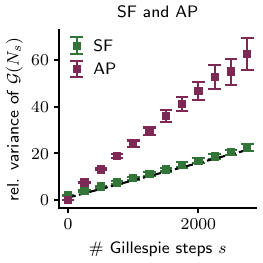}} \hfill
    \subfloat[]{\panel[width=0.48\linewidth]{f}{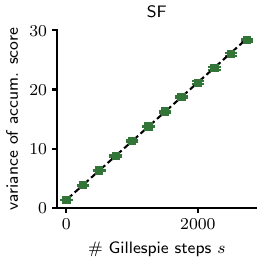}} \\[-3ex]
    \caption{
    Variance scaling of steady-state gradient estimators in the bimolecular association model.
    (a)~The relative variance of the GS-ST estimator increases sharply as the temperature $\tau$ decreases below a critical value.
    (b)~GS-ST variance scales differently with the number of Gillespie steps~$s$ depending on~$\tau$: it converges for large $\tau$, and grows exponentially for small~$\tau$.
    The dashed line shows the analytical estimate of the asymptotic variance for large $\tau$ (Supplementary Material Section \ref{SMsec:sso__variance_gsst}).
    (c)~The Lyapunov exponent indicates parameter regimes with convergent (negative exponent) or divergent variance (positive exponent).
    The temperature at which the Lyapunov exponent changes sign depends on the dissociation rate~$k$.
    (d)~The minimal temperature~$\tau^*$ required to obtain non-diverging variance depends on the parameter regime of~$k$.
    (e)~The variance of the SF and AP gradient estimates grows linearly with~$s$. The dashed line for SF shows the analytical approximation of the variance (Supplementary Material Section~\ref{SMsec:sso__variance_sf}).
    (f)~For SF, the linear variance scaling of the estimated gradient results from the linear variance scaling of the score function; the dashed line shows the analytical estimate.
    Data points show the variance and its uncertainty obtained by computing the variance across $10^4$ stochastic trajectories and repeating this procedure ten times; in panels~(a--b), the error bars indicate the 10th--90th percentile interval instead of the standard deviation.
    Unless stated otherwise, $k = k_{\rm ref} = 5$, $N^{\rm tot}_A = N^{\rm tot}_B = 200$, and $V = 20$.
    }
    \label{fig:bima_var}
\end{figure}

\paragraph*{GS-ST Estimator.}
Fig.~\ref{fig:bima_var}a shows the relative variance of the gradient estimated using GS-ST,
\begin{align}
\frac{\operatorname{Var}\!\left[\mathcal{G}_{\rm GSST}(N_s)\right]}
     {\left({\rm d} \langle N_s \rangle/{\rm d}k\right)^2} \, ,
     \label{eq:gsst_rel_var}
\end{align}
as a function of the estimator-specific temperature $\tau$.
The variance increases significantly as the temperature is lowered, consistent with the behavior observed for the gradient of the loss function (Fig.~\ref{fig:bima_overview_step}d).
Moreover, the critical temperature, below which the variance increases dramatically, depends on the dissociation rate, with more extreme values of $k$ exhibiting a higher critical temperature. 
These features can be understood by analyzing how variances accumulate along the Gillespie trajectories. 
Note that in steady state, ${\rm d} \langle N_s \rangle / {\rm d} k$ and all moments of $N_s$ are independent of $s$, while the relative variance of the estimated gradient (\ref{eq:gsst_rel_var}) is generally not.
Fig.~\ref{fig:bima_var}b shows the relative variance as a function of the number of Gillespie steps $s$. 
At fixed dissociation rate, $k = k_{\rm ref}$, the relative variance converges to a stationary value for sufficiently large~$\tau$.
In contrast, for small~$\tau$, the variance grows exponentially with the number of steps.

To rationalize this behavior, we express the number of complexes at step~$s$ in terms of the number of complexes at the previous step,~$N_s = N_{s-1} + \Delta N_s$. 
Here, $\Delta N_s$~denotes the change in the number of complexes due to the sampled chemical reaction and takes the value~$+1$ for association or~$-1$ for dissociation. 
Using the chain rule to evaluate the gradient ($\Delta N_s$ depends on $k$ explicitly or implicitly via $N_{s-1}$), we can interpret the GS-ST gradient estimator as a linear recursion with random coefficients,
\begin{align*}
    \frac{{\rm d} N_s}{{\rm d} k}
    =
    \left( 1 + \frac{\partial (\Delta N_s)}{\partial N_{s-1}} \right)
    \frac{{\rm d} N_{s-1}}{{\rm d} k}
    + \frac{\partial (\Delta N_s)}{\partial k} \, .
\end{align*}
The variance of ${\rm d} N_s / {\rm d} k$ is thus controlled by the multiplicative factor $1 + \partial (\Delta N_s) / \partial N_{s-1}$.
If its magnitude is typically larger than unity, fluctuations in the gradient are amplified from step to step, leading to exponential growth of the variance.
Conversely, if its magnitude is typically smaller than unity, the variance converges to a finite stationary value.

This intuition can be formalized by introducing the Lyapunov exponent,
\begin{align*}
    \lambda
    =
    \left\langle
    \log \left| 1 + \frac{\partial (\Delta N)}{\partial N} \right|
    \right\rangle \, .
\end{align*}
For a positive Lyapunov exponent, fluctuations in the estimated gradient grow exponentially, while they decay exponentially for a negative Lyapunov exponent (Supplementary Material Section \ref{SMsec:sso__variance_gsst}).
Fig.~\ref{fig:bima_var}c shows that the sign of~$\lambda$ depends sensitively on the temperature~$\tau$, with $\lambda < 0$ at sufficiently high temperature. 
Interestingly, the critical temperature $\tau^{*}$ at which the Lyapunov exponent changes sign depends on the dissociation rate $k$ (Fig.~\ref{fig:bima_var}c).
The critical temperature reaches its minimum close to $k_{\rm ref}$, and increases as $k$ moves away from $k_{\rm ref}$ (Fig.~\ref{fig:bima_var}d). 
This implies that, for fixed temperature $\tau$, the GS-ST estimator can provide low-variance gradient estimates within a window around $k_{\rm ref}$, while entering a high variance regime for very low or high values of $k$. \\

\paragraph*{SF Estimator.}
Unlike the GS-ST estimator, the variance of the SF~estimator grows linearly with the number of Gillespie steps~$s$ (Fig.~\ref{fig:bima_var}e). 
This linear scaling results from an underlying diffusion process in the score function. Note that since
\begin{align*}
    \operatorname{Var}\!\left[\mathcal{G}_{\rm SF}(N_s)\right] = 
    \left\langle\mathcal{G}_{\rm SF}(N_s)^2\right\rangle -
    \left\langle\mathcal{G}_{\rm SF}(N_s)\right\rangle^2 \, ,
\end{align*}
and ${\rm d} \langle N_s \rangle / {\rm d} k$ is constant, 
it suffices to consider the second moment, which using Eqs.~(\ref{eq:SF}) and (\ref{eq:SF__molecule_score}) can be expressed as
{\fontsize{9.5}{11} \selectfont
\begin{align*}
    \left\langle
    \bigl(N_s - \langle N_s \rangle\bigr)^2
    \biggl(
    \sum_{r \leq s} \frac{{\rm d} \log p(\Delta N_r | N_{r-1})}{{\rm d} k}
    \biggl)^2
    \right\rangle \, .
\end{align*}
}\noindent
Since the mean and variance of $N_s$ are constant in steady state, $(N_s-\langle N_s \rangle)^2$~is independent of~$s$, and the entire $s$-dependence arises from the accumulated score term $\sum_{r \leq s} {\rm d}\log p(\Delta N_r \mid N_{r-1}) / {\rm d} k$.
At each Gillespie step, the score function is incremented by a finite contribution, which is positive when a dissociation reaction is sampled and negative when an association reaction occurs, with identical magnitude in both cases (Supplementary Material Section \ref{SMsec:sso__variance_sf}).
Thus, the accumulated score after $s$ steps can be viewed as a sum of binary random variables with zero mean and approximately constant variance, reminiscent of a symmetric random walk.
As a result, its variance grows linearly with $s$ (Fig.~\ref{fig:bima_var}f), implying that $\operatorname{Var}\!\left[\mathcal{G}_{\rm SF}(N_s)\right] \sim s$. 

This linear scaling, resulting from additive accumulation of variance, explains why the SF~estimator performs better than GS-ST in the small $\tau$-regime, where variance accumulates multiplicatively (and hence exponentially). \\

\paragraph*{AP Estimator.}
As for the SF~estimator, the variance of the AP estimator grows linearly with the number of Gillespie steps~$s$, albeit with a higher slope (Fig.~\ref{fig:bima_var}e).
To explain this scaling, we recall the definition of the AP~estimator (Eq.~\ref{eq:AP}) and express its variance as
\begin{align*}
    \operatorname{Var}\!\left[\mathcal{G}_{\rm AP}(N_s)\right]
    \sim
    \left\langle\left(N_s^{\rm alt} - N_s^{\rm prim}\right)^2\, w_{\le s}^2\right\rangle \, ,
\end{align*}
where $N_s^{\rm alt}$ and $N_s^{\rm prim}$ denote the number of complexes in the alternative and primal paths respectively. 
The linear scaling emerges from the interplay between the path difference $N_s^{\rm alt} - N_s^{\rm prim}$ and the accumulated weight~$w_{\le s}$.
For large~$s$, $w_{\le s}^2$ grows quadratically with the number of steps~$s$, whereas the probability that the primal and alternative trajectories differ, $p\left(N_s^{\rm alt} - N_s^{\rm prim} \neq 0\right)$, decreases as $1/s$.
In combination, these two effects cause the variance to grow linearly with $s$ (Supplementary Material Section~\ref{SMsec:sso__variance_ap}).

%%%%%%%%%%%%%%%%%%%%%%%%%%%%%%%%%%%%%%%%%%%%%%%%%%%%%%%%%%%%%%%%%%%%%%%%%%%%%%%%
%%%%%%%%%%%%%%%%%%%%%%%%%% time-dependent estimators %%%%%%%%%%%%%%%%%%%%%%%%%%%
%%%%%%%%%%%%%%%%%%%%%%%%%%%%%%%%%%%%%%%%%%%%%%%%%%%%%%%%%%%%%%%%%%%%%%%%%%%%%%%%
\section{Gradient Estimators for Time-Dependent Observables}

\begin{figure*}
    \begin{tikzpicture}
        \node[anchor=south west, inner sep=0] (img)
        {\includegraphics[width=\linewidth]{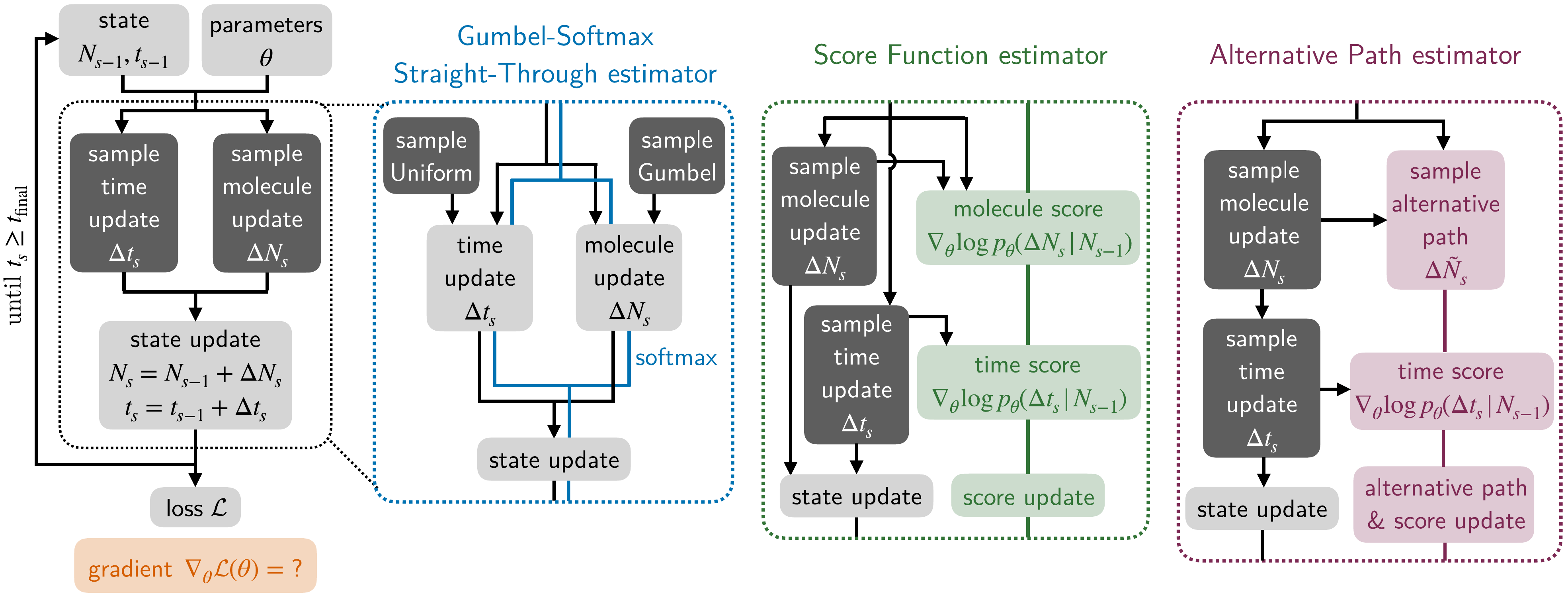}};
        \begin{scope}[x={(img.south east)},y={(img.north west)}]
            \node[anchor=south, xshift=-1pt, yshift=7pt]
            at (0,0.925) {\large\sffamily\bfseries a};
            \node[anchor=south, xshift=-1pt, yshift=7pt]
            at (0.25,0.925) {\large\sffamily\bfseries b};
            \node[anchor=south, xshift=-1pt, yshift=7pt]
            at (0.5,0.925) {\large\sffamily\bfseries c};
            \node[anchor=south, xshift=-1pt, yshift=7pt]
            at (0.75,0.925) {\large\sffamily\bfseries d};
        \end{scope}
    \end{tikzpicture}
    \caption{Illustration of the Gillespie SSA and gradient estimators for time-dependent observables.
    (a)~The Gillespie SSA generates stochastic trajectories by sampling a molecule update~$\Delta N_s$ and a time increment~$\Delta t_s$ at each step~$s$ based on a previous state and input parameters~$\theta$. 
    It is not directly differentiable with respect to~$\theta$ because both $\Delta t_s$ and $\Delta N_s$ are sampled from parameter-dependent probability distributions, and $\Delta N_s$ is a discrete random variable.
    (b)~In the GS-ST~estimator, gradients are obtained by reparameterizing the sampling of $\Delta t_s$ and $\Delta N_s$ using inverse transform sampling for the time and the Gumbel-max trick for the molecule update. 
    For gradient evaluation, the discrete molecule update is approximated by the continuous, differentiable softmax function, enabling automatic differentiation.
    (c)~In the SF~estimator, gradients are obtained by evaluating the product of the simulation output and the score function. 
    At each Gillespie step, score contributions from the molecule and the time update are computed, and accumulated along the trajectory.
    (d)~In the AP~estimator, gradients are obtained by evaluating the weighted difference of an alternative and a primal path. The alternative path~$\Delta \tilde{N}_s$ can either be updated or re-sampled in every Gillespie step. The gradient contribution of the time update $\Delta t_s$ is included through a score term, analogous to the SF~estimator.
    }
    \label{fig:estimator_illustration}
\end{figure*}

So far, we have focused on estimating gradients of trajectories with a fixed number of Gillespie steps.
While this setting is sufficient for steady-state observables, many applications require observables evaluated at a fixed physical time~$t_{\rm final}$.
In the Gillespie SSA, this time dependence enters through the stochastic waiting times~$\Delta t_s$ sampled between successive reaction events, and the dynamics are simulated until $t_s\ge t_{\rm final}$ (Fig.~\ref{fig:estimator_illustration}a).
Since the waiting times depend on the system parameters, they can provide a non-vanishing contribution to the gradient. We next turn to the specific form of this contribution, which depends on the estimator. 

\subsection{Adapting Gradient Estimators to Time-Dependent Observables}

\paragraph*{GS-ST Estimator.} 
In the time-dependent case, each Gillespie step $s$ samples not only the change in molecule count $\Delta N_s$ (left branch in Fig.~\ref{fig:estimator_illustration}b) but also the waiting time $\Delta t_s$ until the next reaction (right branch in Fig.~\ref{fig:estimator_illustration}b).
The change in molecule number $\Delta N_s$ and its gradient are evaluated as described for steady-state observables (see previous section).
The waiting time is sampled via inverse transform sampling, \looseness=-1 $\Delta t_s = - \ln u'_s / a_s^{\rm tot}$, with $u'_s \sim \operatorname{Uniform}(0,1)$. 
Since $u'_s$ is independent of the parameters, $\nabla_\theta \Delta t_s$ is accessible through standard automatic differentiation.
The molecule count at time $t_{\rm final}$ can be written as
\begin{align*}
    N(t_{\rm final}) = N_0 + \sum_s \Delta N_s\, \Theta(t_{\rm final} - t_s) \, ,
\end{align*}
where the Heaviside step function ensures that only reaction events with $t_s \leq t_{\rm final}$ contribute. 
However, this sharp temporal cutoff prevents direct gradient evaluation. 
To obtain a differentiable approximation, we replace the Heaviside function by a sigmoid function,
\begin{align}
    N(t_{\rm final}) \approx N_0 + \sum_s \Delta N_s \,
    \sigma\!\left( \frac{t_{\rm final} - t_s}{\tau_{\rm time}} \right) \,,
    \label{eq:gsst_mol_count_relaxed_time}
\end{align}
where $\tau_{\rm time}$ sets the characteristic timescale of the temporal cutoff. The resulting expression defines a differentiable surrogate of $N(t_{\rm final})$. Since the sigmoid function is differentiable, and both $\nabla_\theta \Delta N_s$ and $\nabla_\theta \Delta t_s$ are accessible through automatic differentiation, the gradient $\nabla_\theta N(t_{\rm final})$ can be evaluated within the GS-ST estimator. \looseness=0 \\

\paragraph*{SF Estimator.}
Each time-dependent Gillespie trajectory is fully specified by the sequence of sampled reaction channels and associated waiting times.
The distribution of trajectories is characterized by the joint distribution $p(\Delta N_{\le s}, \Delta t_{\le s})$, with $s$ denoting the number of Gillespie steps required to reach $t_{\rm final}$.
The corresponding score function can be decomposed as
\begin{align}
    \nabla_\theta \log p(\Delta N_{\le s},\Delta t_{\le s}) &= \nabla_\theta \log p(\Delta N_{\le s}) \notag \\
    &\quad + \nabla_\theta \log p(\Delta t_{\le s} \mid \Delta N_{\le s}) \,.
    \label{eq:SF__joint_score_function}
\end{align}
The first term, referred to as the molecule score in Fig.~\ref{fig:estimator_illustration}c, is the score contribution associated with the sequence of sampled molecule updates, as discussed for steady-state observables (see previous section, Eq.~(\ref{eq:SF__molecule_score})).
The second term in Eq.~(\ref{eq:SF__joint_score_function}), which we refer to as the time score, accounts for the stochasticity of the waiting times. 
Since the waiting times are mutually independent and each $\Delta t_s$ depends only on the preceding reaction history $\Delta N_{\leq s}$, the time score is additive,
\begin{align*}
    \nabla_\theta \log p(\Delta t_{\le s}\mid \Delta N_{\le s})
    =
    \sum_{r \leq s} \nabla_\theta \log p(\Delta t_r \mid N_{r-1}) \, .
\end{align*}
This expression can be evaluated analytically: each $\Delta t_s$ is drawn from an exponential distribution with a rate set by the total propensity $a_{s-1}^{\rm tot}$, implying
\begin{align*}
    \nabla_\theta \log p(\Delta t_s \mid N_{s-1})
    =
    \left( \frac{1}{a_{s-1}^{\rm tot}} - \Delta t_s \right) \nabla_\theta \, a_{s-1}^{\rm tot} \, .
\end{align*}
Including both the molecule score and the time score yields an unbiased SF estimator for gradients at fixed physical time. \\

\paragraph*{AP Estimator.}
In the AP estimator, the gradient contribution due to the molecule update is obtained by evaluating the weighted difference between the primal and alternative paths (upper half in Fig.~\ref{fig:estimator_illustration}d, see previous section). 
Importantly, for time-dependent observables, alternative and primal path must be evaluated at the same observation time.
Since the two paths evolve with distinct system times, $t^{\rm prim}$ and $t^{\rm alt}$, this common time may be reached after different numbers of Gillespie steps in the two paths.
In addition, the gradient contribution of the waiting times is accounted for by including a time score, in analogy with the SF estimator (lower half in Fig.~\ref{fig:estimator_illustration}d). 
The full gradient estimate is obtained by summing the molecule-update and waiting-time gradient contributions.

\subsection{Example: Bimolecular Association}

\begin{figure}
    \vspace{-10pt}
    \subfloat[]{\panel[width=0.48\linewidth]{a}{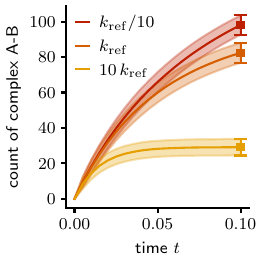}} \hfill
    \subfloat[]{\panel[width=0.48\linewidth]{b}{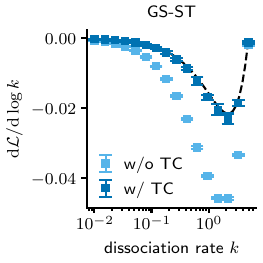}} \\[-7ex]
    \subfloat[]{\panel[width=0.48\linewidth]{c}{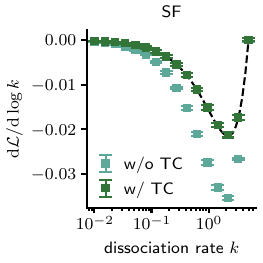}} \hfill
    \subfloat[]{\panel[width=0.48\linewidth]{d}{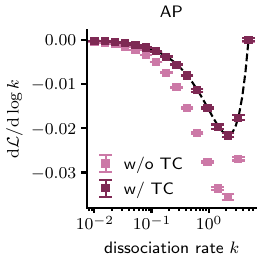}} \\[-3ex]
    \caption{
    Bimolecular association model evaluated at fixed time.
    (a) Starting from an initial configuration without bimolecular complexes, the system reaches a nonzero number of complexes by the evaluation time $t_{\rm final}=0.1$.
    Solid lines and shaded areas show the mean and standard deviation of $10^4$ trajectories.
    (b--d)~Gradient of the loss function evaluated at $t_{\rm final}=0.1$ using the GS-ST~(b), SF~(c), and AP~(d) estimators. For GS-ST, $\tau = 0.03$ and $\tau_{\rm time} = 2.5 \cdot 10^{-5}$.
    The estimated gradients agree with the analytical reference (dashed lines) only when the 
    contribution of the waiting times to the gradient is explicitly taken into account (with time contribution (TC), dark data points); otherwise the gradient is overestimated (without TC, light data points). 
    Data points show mean values and uncertainties by averaging over ten batches, each consisting of $10^4$ trajectories.
    In all panels, $k_{\rm ref} = 5$, $N^{\rm tot}_A = N^{\rm tot}_B = 200$, and $V = 20$.
    }
    \label{fig:bima_overview_time}
\end{figure}

To test the time-dependent gradient estimators, we revisit the bimolecular association model, now evaluating the simulation output after a fixed time $t_{\rm final} = 0.1$ rather than after a fixed number of Gillespie steps.
Fig.~\ref{fig:bima_overview_time}a shows the time evolution of the count of bimolecular complexes: 
starting from zero, the number of complexes increases towards its steady-state count, with the relaxation timescale depending on the dissociation rate $k$.
For high values of $k$, the system reaches its steady state by $t_{\rm final}$, while the steady state is not reached for low $k$.

We quantify differences between the trajectories at the evaluation time using the time-dependent loss function \looseness=-1
\begin{align*}
    \mathcal{L}(k, t) = \left( \frac{\langle N(k) \rangle_t - \langle N(k_{\rm ref}) \rangle_t}{\langle N(k_{\rm ref}) \rangle_t}\right)^2 \,.
\end{align*}
When gradients of this loss function are computed using the steady-state estimators without modification, none of the three estimators reproduces the expected gradient (light data points in Fig.~\ref{fig:bima_overview_time}b--d).
This deviation arises because the physical time elapsed after $s$ Gillespie steps is itself a stochastic quantity: the waiting times are drawn from parameter-dependent distributions, so changes in $k$ affect not only which reactions are selected, but also the waiting time between them. Once the estimators are adapted to account for this additional gradient contribution (as described above), all three estimators recover the correct gradient (dark data points in Fig.~\ref{fig:bima_overview_time}b--d).

\subsection{Variance Scaling of Gradient Estimators}

\begin{figure}
    \vspace{-10pt}
    \subfloat[]{\panel[width=0.48\linewidth]{a}{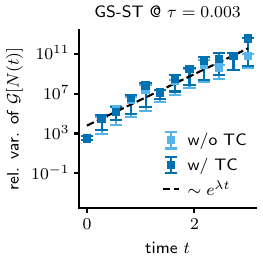}} \hfill
    \subfloat[]{\panel[width=0.48\linewidth]{b}{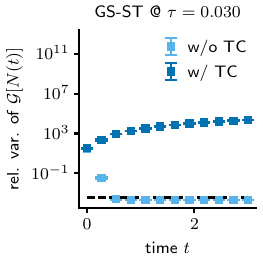}} \\[-7ex]
    \subfloat[]{\panel[width=0.48\linewidth]{c}{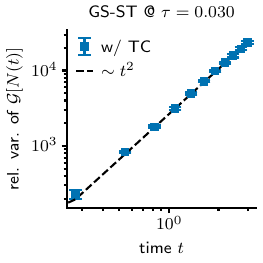}} \hfill
    \subfloat[]{\panel[width=0.48\linewidth]{d}{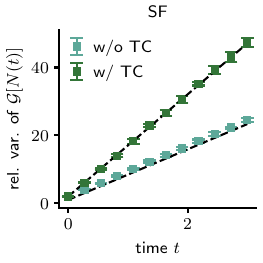}} \\[-3ex]
    \caption{
    Variance scaling of time-dependent gradient estimators in the bimolecular association model.
    (a)~For small $\tau$, the relative gradient variance of the GS-ST estimator increases exponentially with time, irrespective of whether the waiting-time contribution is accounted for (with TC) or not (without TC).
    (b)~For large $\tau$, including the waiting-time contribution increases the GS-ST variance relative to the time-independent case; the dashed line shows the analytical approximation of the time-independent asymptotic variance (Supplementary Material Section~\ref{SMsec:sso__variance_gsst}).
    (c)~The variance of the time-dependent GS-ST estimator scales quadratically with $t$ for large $\tau$.
    (d)~The variance of the SF estimator scales linearly with $t$ (dark green), analogously to the time-independent case (light green).
    Data points show the variance and its uncertainty obtained by computing the variance across $10^4$ stochastic trajectories and repeating this procedure ten times; in panels~(a--c), the error bars indicate the 10th--90th percentile interval instead of the standard deviation.
    In all panels, $\tau_{\rm time} = 2.5 \cdot 10^{-5}$, $k_{\rm ref} =5$, $N^{\rm tot}_A = N^{\rm tot}_B = 200$, and $V = 20$.
    }
    \label{fig:bima_var_time}
\end{figure}

Including the waiting time contribution in the gradient estimators can affect the gradient variance, since the waiting times introduce an additional source of stochasticity.
As for the variance scaling analysis of time-independent observables, we investigate how the gradient variance scales with the evaluation time $t$ for simulations initialized in steady state. 
This ensures that the observed variance reflects the intrinsic properties of the estimators, rather than the relaxation dynamics of the bimolecular association process. \\

\paragraph*{GS-ST Estimator.} 
The effect of including the waiting-time contribution to the gradient strongly depends on the temperature $\tau$. 
For $\tau$ below the critical temperature, the waiting-time contribution increases the relative variance 
\begin{align*}
    \frac{\operatorname{Var}[\mathcal{G}_{\rm GSST}[N(t)]]}{({\rm d} \langle N(t) \rangle / {\rm d} k)^2}
\end{align*}
only marginally (Fig.~\ref{fig:bima_var_time}a). 
In this regime, the variance already grows exponentially, due to the molecule-update contribution, rendering the additional waiting-time contribution negligible in comparison. 
Above the critical temperature, the variance of the time-independent gradient estimator converges to a finite asymptotic value (light blue data points in Fig.~\ref{fig:bima_var_time}b). Including the waiting-time contribution causes the variance to grow quadratically with time $t$ (dark blue points in Fig.~\ref{fig:bima_var_time}b--c).

The quadratic scaling can be understood by differentiating the molecule count (\ref{eq:gsst_mol_count_relaxed_time}) with respect to $k$,
\begin{align*}
    \frac{{\rm d} N(t)}{{\rm d} k} = \, & \frac{{\rm d} N_0}{{\rm d} k} + \sum_s \frac{{\rm d} (\Delta N_s)}{{\rm d} k} \sigma\left( \frac{t - t_s}{\tau_{\rm time}} \right) \\
    &+ \sum_s \Delta N_s \frac{\partial \sigma\left( \frac{t-t_s}{\tau_{\rm time}}\right)}{\partial t_s} \frac{{\rm d} t_s}{{\rm d} k} \, .
\end{align*}
Here, the first two terms account for the molecule-update contribution, while the last term adds the waiting-time contribution. 
The variance of the gradient estimate is 
\begin{align*}
    \operatorname{Var}\left[\frac{{\rm d} N(t)}{{\rm d} k} \right] 
    = \left\langle \left( \frac{{\rm d} N(t)}{{\rm d} k} \right)^2 \right\rangle
    - \left\langle \frac{{\rm d} N(t)}{{\rm d} k}\right\rangle^2 \, .
\end{align*}
Since the system is in steady state, the expectation value $\langle {\rm d} N(t) / {\rm d} k \rangle$ is time-independent, implying that this term does not contribute to the scaling of the variance with time.
Neglecting covariances, we approximate
{\fontsize{9.5}{11} \selectfont
\begin{align*}
    \left\langle \left(\frac{{\rm d} N(t)}{{\rm d} k} \right)^2 \right\rangle \approx \sum_s \left\langle \Delta N_s^2 \left( \frac{{\rm d} \sigma\left( \frac{t-t_s}{\tau_{\rm time}} \right)}{{\rm d} t_s} \right)^2 \left(\frac{{\rm d} t_s}{{\rm d} k} \right)^2\right\rangle \, .
\end{align*}
}\noindent
In steady-state, the factor $\Delta N_s^2$ is independent of time.
Therefore, the time-dependence of the variance arises from the remaining factors, $\operatorname{Var}[{\rm d} N(t) / {\rm d} k] \sim \sum_s \left\langle ({\rm d} \sigma / {\rm d} t_s)^2 \, ({\rm d} t_s / {\rm d} k)^2 \right\rangle$. 
The derivative of the sigmoid function acts as a localized kernel around $t_s = t$, ensuring that only terms within a time window of width proportional to $\tau_{\rm time}$ around $t$ contribute in the sum over $s$, such that $\operatorname{Var}[{\rm d} N(t) / {\rm d} k] \sim \langle({\rm d} t / {\rm d} k)^2 \rangle$. 
This expectation value scales as $\langle ({\rm d} t / {\rm d} k)^2 \rangle \sim t^2$, explaining the quadratic scaling of the gradient variance (see Supplementary Material Section \ref{SMsec:tdo__variance_gsst} for the full derivation). 

The prefactor of this power law is proportional to the inverse of the characteristic timescale $\tau_{\rm time}$. 
A softer temporal cutoff (higher $\tau_{\rm time}$) therefore reduces the variance of the estimated gradients.
However, increasing $\tau_{\rm time}$ also increases the bias, and sufficiently unbiased samples are attained only for characteristic timescales shorter than the typical inverse propensity, $\tau_{\rm time} < \langle a^{\rm tot} \rangle^{-1}$ (see Supplementary Material Section \ref{SMsec:tdo__variance_gsst} for a discussion of this bias-variance trade-off). \\

\paragraph*{SF Estimator.}
In the GS-ST estimator, including the time-dependence changes the scaling behavior of the gradient variance for high temperature (from constant to $t^2$). 
In contrast, the variance of the SF estimator scales linearly with time in both cases, with the inclusion of the time-dependence merely affecting the slope (Fig.~\ref{fig:bima_var_time}d).
To rationalize this, we follow the same argument as for the time-independent SF estimator. 
Since the system is in steady state, the molecule count $N(t)$ fluctuates around a constant mean, and the variance of the estimator is set by the variance of the accumulated score function. 
For time-dependent gradients, the score function comprises the molecule score and the time score,
\begin{align*}
    \operatorname{Var}[\mathcal{G}_{\rm SF}[N(t)]] \sim \operatorname{Var}\!\bigg[ \sum_{r \leq s} \bigg( &\frac{{\rm d} \log p(\Delta N_r \mid N_{r-1})}{{\rm d} k} \\
    &+ \frac{{\rm d} \log p(\Delta t_r \mid N_{r-1})}{{\rm d} k} \bigg) \bigg] \, .
\end{align*}
As seen above, the variance of the molecule score grows linearly with the number of Gillespie steps $s$, leading to linear scaling with $t$, because the average time increment per step is constant in steady state.
Similarly, the time score is a sum of (approximately) independent per-step contributions, and its variance therefore also grows linearly with $t$. 
The total variance of the SF estimator, corresponding to the sum of both variances, thus scales linearly with $t$, with a slope corresponding to the sum of the per-step molecule and time score variances (see Supplementary Material Section~\ref{SMsec:tdo__variance_sf} for the derivation). \\

\paragraph*{AP estimator.}
The variance of the AP estimator for time-dependent observables initially grows linearly with time, but eventually crosses over to superlinear growth (Supplementary Fig.~\ref{SMfig:tdo__ap_var_scaling}).
The initial regime mirrors the linear variance scaling observed for the time-independent AP estimator (Fig.~\ref{fig:bima_var}e), since the number of Gillespie steps $s$ is proportional to the simulation time $t$ on average.
The origin of the crossover is more subtle.
In the time-independent case, the variance grows linearly with $s$ even for high $s$: although the second moment of the accumulated weight grows quadratically with $s$, the probability of obtaining a non-zero difference between alternative and primal path decays as $1/s$. These two effects combine to give an overall linear scaling (see also Supplementary Material Section~\ref{SMsec:sso__variance_ap}).
For time-dependent observables, the accumulated weight also grows quadratically with time.
However, the probability of a non-zero path difference no longer decays to zero, but approaches a finite constant.
As a result, the cancellation that yields linear scaling in the time-independent case is lost, causing the variance to grow superlinearly with time (Supplementary Material Section~\ref{SMsec:tdo__variance_ap}). \looseness=-1

%%%%%%%%%%%%%%%%%%%%%%%%%%%%%%%%%%%%%%%%%%%%%%%%%%%%%%%%%%%%%%%%%%%%%%%%%%%%%%%%
%%%%%%%%%%%%%%%%%%%%%%%%%%%%%% parameter inference %%%%%%%%%%%%%%%%%%%%%%%%%%%%%
%%%%%%%%%%%%%%%%%%%%%%%%%%%%%%%%%%%%%%%%%%%%%%%%%%%%%%%%%%%%%%%%%%%%%%%%%%%%%%%%
\section{Gradient-based Parameter Inference in Time-Dependent Systems}
Up to this point, we have shown that the GS-ST, SF and AP estimators can be used to evaluate gradients of steady-state and time-dependent observables, and have characterized their properties using the bimolecular association process as an example.
We now apply the estimators to parameter inference in a chemical reaction network that does not converge to a steady state, but instead exhibits self-sustained oscillations.
Due to the oscillations, parameter inference must take the full temporal structure of the trajectories into account, providing a challenging task for the time-dependent gradient estimators.

\subsection{Example: Repressilator}
As a benchmark example, we consider the repressilator~\citep{elowitz2000syntheticoscillatory, potvin2016synchronous}, a network of three protein species that repress each other's production in a cyclic manner (Fig.~\ref{fig:repr_inf}a). 
Specifically, the production rate of protein species $i$ is inhibited by the abundance of protein species~$i-1$. 
This repression is described by a Hill function, such that the production propensity of species~$i$ is given by
\begin{align*}
    a_{\mathrm{prod},\,i}
    =
    \frac{k_{\rm p} V}
    {1 + \left( \frac{N_{i-1}}{K_{\rm d} V} \right)^h } \, .
\end{align*}
Here, $k_{\rm p}$ denotes the basal production rate, $K_{\rm d}$~the dissociation constant of the repressor protein to the regulatory site, $h$ the Hill coefficient, and $V$~the system volume. 
In addition to production, all protein species undergo degradation, which we model as a first-order process with propensity $a_{\mathrm{deg},\,i} = k_{\rm d} N_i$. 
For simplicity, we set~$k_{\rm d}=1$, such that the intrinsic timescale of the system is set by the protein lifetime, and keep the volume~$V$ fixed throughout.

Analysis of the corresponding deterministic chemical rate equations shows that sustained oscillations arise when the ratio $k_{\rm p}/K_{\rm d}$ is sufficiently large~\citep{loinger2007stochastic}.
The amplitude and period of these oscillations depend on the values of $k_{\rm p}$ and $K_{\rm d}$, but are largely insensitive to the Hill coefficient, provided that it is large enough to permit oscillations (Appendix~\ref{APsec:repr_mf}). 
For this reason, we fix the Hill coefficient to $h=3$ and aim at inferring only the two-dimensional parameter vector $k = (k_{\rm p}, K_{\rm d})$. 
Stochastic simulations with the Gillespie algorithm confirm that the system indeed exhibits oscillatory dynamics in this regime, even in the presence of intrinsic noise (Fig.~\ref{fig:repr_inf}b).

As in the bimolecular association example, we consider the inverse problem of inferring a reference parameter set~$k_{\rm ref}$ from stochastic trajectories generated at that parameter value. To compare trajectories obtained at a test parameter set~$k$ with those at $k_{\rm ref}$, we define the loss function
{\fontsize{9.5}{11}\selectfont 
\begin{align*}
    \mathcal{L}(k)
    =
    \frac{1}{3\, n_{\rm times}}
    \sum_{i,j}
    \left[
    \log \langle N_i(k) \rangle_{t_j}
    -
    \log \langle N_i(k_{\rm ref}) \rangle_{t_j}
    \right]^2 \, .
\end{align*}
}\noindent
The loss measures the squared deviation between the logarithms of the average copy numbers of the three protein species, evaluated at a discrete set of time points; in practice, we choose~$n_{\rm times} = 10$ time points.
Defining the loss function in terms of the logarithms of the copy numbers ensures that relative deviations contribute equally to the loss, independent of the absolute molecule number.
For each reference parameter set, the observation time window is chosen to span one oscillation period, ensuring that the loss captures the full temporal structure of the dynamics.

\subsection{Parameter Inference}
\begin{figure}
    \vspace{-10pt}
    \begin{minipage}[t]{0.36\linewidth}
    %
  % #1 = optional includegraphics options (e.g., width=0.45\linewidth)
  % #2 = panel label, e.g. a, b, c
  % #3 = filename
  \begin{tikzpicture}[baseline=(img.north)]
    \node[anchor=south west, inner sep=0] (img)
      {\resizebox{0.99\linewidth}{!}{\begin{tikzpicture}[
    gene/.style={circle, draw, thick, minimum size=0.5cm, font=\bfseries},
    inhibit/.style={-{Bar[length=0pt,width=8pt]}, thick, shorten >=2mm, shorten <=2mm},
]
\definecolor{gene1col}{HTML}{bd1f01}
\definecolor{gene2col}{HTML}{D55E00}
\definecolor{gene3col}{HTML}{E69F00}
% Canvas size
\def\W{3cm}
\def\H{4.19cm}

\path[use as bounding box] (0,0) rectangle (\W,\H);

% Center everything
\begin{scope}[shift={(\W/2,\H/2 - 0.1cm)}]

\node[gene, fill=gene1col, text=gene1col, draw=gene1col, fill opacity=0.3, text opacity=1.] (1) at (90:1.3cm)  {1};
\node[gene, fill=gene2col, text=gene2col, draw=gene2col, fill opacity=0.3, text opacity=1.] (2) at (210:1.3cm) {2};
\node[gene, fill=gene3col, text=gene3col, draw=gene3col, fill opacity=0.3, text opacity=1.] (3) at (330:1.3cm) {3};

\draw[inhibit] (1) -- (2);
\draw[inhibit] (2) -- (3);
\draw[inhibit] (3) -- (1);

\end{scope}
\end{tikzpicture}}};
    \node[anchor=base west, xshift=-1pt, yshift=0pt]
      at (img.north west) {\large \sf \textbf{a}};
  \end{tikzpicture}%

    \end{minipage}
    \hfill
    \begin{minipage}[t]{0.62\linewidth}
    \panel[width=0.99\linewidth]{b}{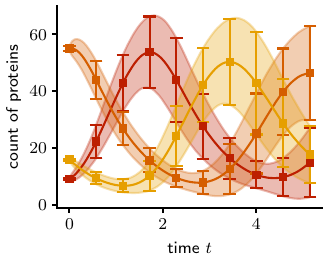}
    \end{minipage}
    \\[-3ex]
    \subfloat[]{\panel[width=0.49\linewidth]{c}{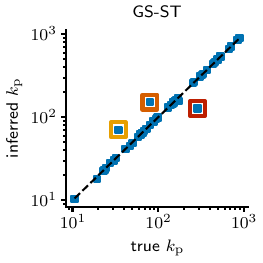}} \hfill
    \subfloat[]{\panel[width=0.49\linewidth]{d}{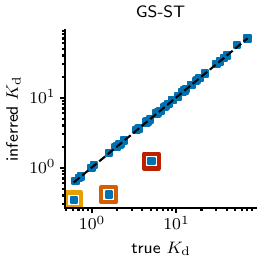}} \\[-7ex]
    \subfloat[]{\panel[width=0.49\linewidth]{e}{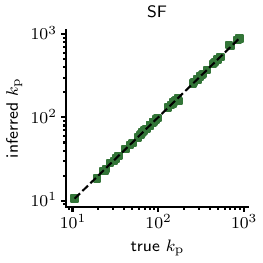}} \hfill
    \subfloat[]{\panel[width=0.49\linewidth]{f}{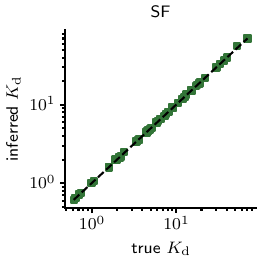}} \\[-7ex]
    \subfloat[]{\panel[width=0.49\linewidth]{g}{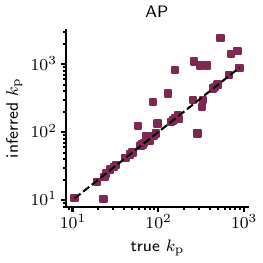}} \hfill
    \subfloat[]{\panel[width=0.49\linewidth]{h}{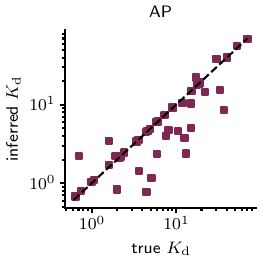}} \\[-4ex]
    \caption{
    Parameter inference in the repressilator model.
    (a)~Three protein species are produced and degraded, and mutually repress each other’s production.
    (b)~In certain parameter regimes, the system exhibits self-sustained oscillations under intrinsic noise. Mean and standard deviation are obtained from $10^4$ Gillespie trajectories with $k_{\rm p}=100$, $K_{\rm d}=10$, $h=3$, and $V=1$.
    (c--h)~The production rate constant~$k_{\rm p}$ and dissociation constant~$K_{\rm d}$ are inferred from trajectory data by minimizing the loss function ($h=3$, $V=1$ fixed).
    Results are shown for 50~optimization runs, corresponding to different reference parameter sets and different initializations of the optimization.
    Panels (c--d), (e--f), and (g--h) compare inferred and reference parameter values obtained using GS-ST, SF, and AP, respectively. 
    In each optimization step, the gradient of the loss is estimated from 1000~stochastic trajectories; for GS-ST, we use $\tau = 0.3$ and $\tau_{\rm time} = 0.05$. 
    }
    \label{fig:repr_inf}
\end{figure}

\begin{figure*}
    \subfloat[]{\panel[width=0.32\linewidth]{a}{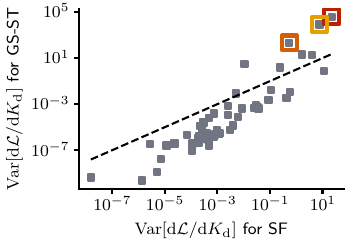}}\hfill
    \subfloat[]{\panel[width=0.32\linewidth]{b}{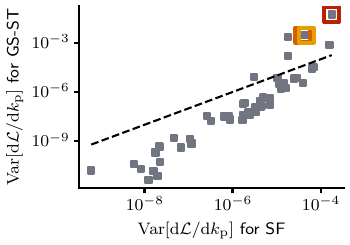}}\hfill
    \subfloat[]{\panel[width=0.32\linewidth]{c}{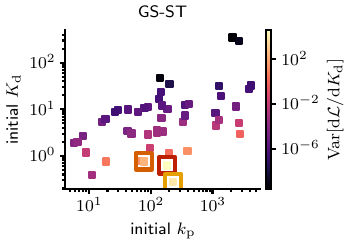}} \\[-5ex]
    \subfloat[]{\panel[width=0.32\linewidth]{d}{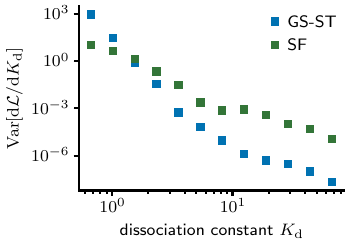}}\hfill
    \subfloat[]{\panel[width=0.32\linewidth]{e}{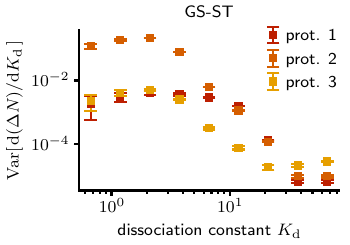}}\hfill
    \subfloat[]{\panel[width=0.32\linewidth]{f}{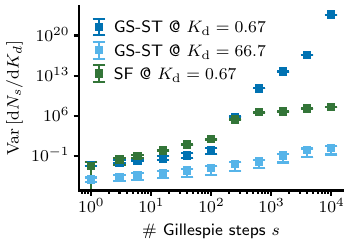}}\\[-5ex]
    \caption{
    Divergent gradient variance explains failures of parameter inference with the GS-ST estimator.
    (a--b)~For the three parameter sets where GS-ST fails to converge to the true optimum, the gradients with respect to $k_{\rm p}$ and $K_{\rm d}$ exhibit high variance.
    (c)~The GS-ST gradient variance depends sensitively on the system parameters; in particular, the variance is larger for stronger binding affinity (lower $K_{\rm d}$).
    In panels (a--c), the variances are evaluated at the parameter values used to initialize the optimization.
    (d)~For GS-ST and SF, the variance of the loss gradients increases sharply with binding affinity, although the dependence on $K_{\rm d}$ is weaker for SF.
    (e)~The variance of the gradient of the change in molecule count in a single Gillespie step $\Delta N$ increases as $K_{\rm d}$ decreases, with comparable sensitivity to $K_{\rm d}$ for GS-ST and SF (Supplementary Fig.~\ref{SMfig:repr__variance_sf}).
    (f)~For GS-ST, the variance of the gradients of the molecule counts $N_s$ diverges at low $K_{\rm d}$ (dark blue), but increases approximately linearly with $s$ for high $K_{\rm d}$ (light blue).
    By contrast, SF does not exhibit divergent variance even at strong binding affinity and instead scales approximately linearly with $s$ for sufficiently large $s$ (dark green).
    For panels (d--f), variance data are obtained at fixed $k_{\rm ref}=(200,6.67)$ and $k_{\rm p}=200$, while varying $K_{\rm d}$.
    In all panels, variances of the loss gradients are computed by averaging gradients over $10^3$~independent trajectories and evaluating the variance across 50~repetitions.
    Variances of molecule counts are computed by first evaluating the variance across $10^4$~trajectories and then averaging over 100~repetitions.
    }
    \label{fig:repr_var}
\end{figure*}

Using the loss function defined above, we perform parameter inference by searching for the parameter vector~$k$ at which the loss attains its minimum. 
To this end, we employ stochastic gradient descent~(SGD). 
Starting from an initial parameter set $k$, we estimate the gradient of the loss function using one of the three gradient estimators introduced above, and then update the parameters by moving them in the direction of the negative gradient. 
Since both parameters vary over several orders of magnitude, we perform the optimization in log-parameter space,
\begin{align*}
    \log k_{\rm p} &\mapsto \log k_{\rm p} - \alpha \frac{{\rm d} \mathcal{L}}{{\rm d} \log k_{\rm p}} \, , \\
    \log K_{\rm d} &\mapsto \log K_{\rm d} - \alpha \frac{{\rm d} \mathcal{L}}{{\rm d} \log K_{\rm d}} \, .
\end{align*}
with learning rate~$\alpha=0.1$, which controls the step size of the optimization.
The parameters are updated iteratively as long as the decrease in the loss remains sufficiently well resolved relative to its fluctuations. 
To quantify this, we monitor the signal-to-noise ratio of the loss change,
\begin{align*}
    \mathrm{SNR}(\Delta \mathcal{L})
    =
    \frac{\mathrm{median}(\Delta \mathcal{L})}
         {\mathrm{median}\!\left(
         \bigl| \Delta \mathcal{L} - \mathrm{median}(\Delta \mathcal{L}) \bigr|
         \right)} \, .
\end{align*}
Here, $\Delta \mathcal{L}$~denotes the change in the loss function per optimization step, and both medians are evaluated over the preceding 50~optimization steps. 
We terminate the optimization once the signal-to-noise ratio drops below~$0.01$ three times.

To test the parameter inference procedure, we sample 50~reference parameter sets at random, drawing $k_{\rm p}^{\rm ref}$ from a log-uniform distribution between 10 and 1000 and $K_{\rm d}^{\rm ref}$ from a log-uniform distribution between $k_{\rm p}^{\rm ref}/75$ and $k_{\rm p}^{\rm ref}/3$. 
The upper bound $K_{\rm d} < k_{\rm p}/3$ ensures that the system falls in the regime of self-sustained oscillations. 
For each reference parameter set, we perform a separate inference run starting from randomly chosen initial parameters sampled log-uniformly from the intervals $(k_{\rm p}^{\rm ref} / 10, ~ 10 \, k_{\rm p}^{\rm ref})$ and $(K_{\rm d}^{\rm ref} / 10, ~ 10 \, K_{\rm d}^{\rm ref})$.

The results of the parameter inference procedure are summarized in Fig.~\ref{fig:repr_inf}c--h, where the inferred parameters are plotted against the corresponding reference parameters. 
Each inference run uses 1000~stochastic trajectories to estimate the loss function and its gradient. 
Overall, parameter inference based on the GS-ST and SF~estimators recovers the reference parameters well, whereas the AP~estimator performs markedly worse. 

To explain the poor performance of the AP~estimator, we compare the variance of the gradients estimated with AP and SF.
Throughout the following analysis, gradient variance is evaluated by averaging the gradient over 1000 independent stochastic trajectories and computing the variance across 50 repetitions of this procedure for each parameter set. 
Across all parameter sets, AP exhibits on average about 50-fold larger gradient variances, consistent with its poorer convergence in the inference procedure (Supplementary Material Section \ref{SMsec:repr__variance_ap}). 
Although inference with AP can be improved by increasing the number of trajectories used for gradient evaluation, its high variance renders the AP~estimator unattractive for parameter inference in the repressilator system.

Among GS-ST and SF, the SF~estimator shows slightly more robust convergence: 
it recovers the reference parameters for all 50~tested cases, whereas GS-ST fails to converge in three cases (highlighted by colored boxes in Fig.~\ref{fig:repr_inf}c--d). 
To understand why GS-ST performs worse than SF in these cases, we analyze the variance of the estimated gradients at the parameter values used to initialize the optimization. 
For the three parameter sets, we find that the GS-ST estimated gradients exhibit markedly higher variance than those obtained with SF, whereas across most of the remaining parameter sets, GS-ST yields lower-variance gradient estimates than SF (Fig.~\ref{fig:repr_var}a--b).

Interestingly, the variance of the gradients depends strongly on the system parameters: 
increasing the binding affinity (i.e., decreasing $K_{\rm d}$) leads to a substantial increase in variance. 
Consistent with this trend, all problematic optimizations are initialized in a regime of high binding affinity (Fig.~\ref{fig:repr_var}c). 
To analyze the $K_{\rm d}$-dependence of the variance more quantitatively, we fix the reference parameter set to $k_{\rm ref} = (200,6.67)$ and vary only the parameter $K_{\rm d}$ while keeping $k_{\rm p}=k_{\rm p}^{\rm ref}$ constant (Fig.~\ref{fig:repr_var}d). 
The variance of the gradient ${\rm d}\mathcal{L}/{\rm d}K_{\rm d}$ estimated with GS-ST increases dramatically as $K_{\rm d}$ is reduced, growing by more than ten orders of magnitude as $K_{\rm d}$ decreases by only two orders of magnitude (blue data points in Fig.~\ref{fig:repr_var}d). 
By comparison, the variance of the SF~estimator shows a much weaker dependence on $K_{\rm d}$: although it is larger than that of GS-ST at high $K_{\rm d}$, it increases more slowly and eventually remains well below the GS-ST variance in the regime of small $K_{\rm d}$ (green data points in Fig.~\ref{fig:repr_var}d).

The pronounced $K_{\rm d}$ dependence of the gradient variance arises from two compounding contributions: 
(i)~the dependence of the single-step variance on $K_{\rm d}$, and 
(ii)~the $K_{\rm d}$-dependent variance propagation along the chain of Gillespie steps. 
We find that for \emph{both} estimators, the variance associated with an individual step increases substantially as $K_{\rm d}$ is reduced (Fig.~\ref{fig:repr_var}e for GS-ST and Supplementary Fig.~\ref{SMfig:repr__variance_sf} for SF). 
Consequently, this increase alone does not explain the difference between the two estimators. 
The key difference is that, for sufficiently small $K_{\rm d}$, the GS-ST estimator enters a regime in which gradient variances accumulate multiplicatively along the trajectory, causing the total variance to grow rapidly with trajectory length (dark blue data points in Fig.~\ref{fig:repr_var}f).
By contrast, the SF estimator does not show this behavior: for large~$s$, its variance grows only linearly with the number of Gillespie steps (green data points in Fig.~\ref{fig:repr_var}f).
\looseness=-1
A natural strategy to mitigate the large variance at small $K_{\rm d}$ is to increase the temperature $\tau$ of the GS-ST estimator. 
Indeed, larger temperature~$\tau$ restores convergence for some of the difficult parameter sets discussed above. 
However, this comes at the cost of introducing a new set of non-converging inference runs. 
As shown in Appendix~\ref{APsec:repr_bias_gsst}, these failures are not caused by large variance, but by a bias in the estimated gradients, which causes the optimization to miss the true optimum. 
%The repressilator system illustrates that already a simple example with two parameters can present a challenging benchmark for parameter inference and robust gradient estimators 

%%%%%%%%%%%%%%%%%%%%%%%%%%%%%%%%%%%%%%%%%%%%%%%%%%%%%%%%%%%%%%%%%%%%%%%%%%%%%%%%
%%%%%%%%%%%%%%%%%%%%%%%%%%%%%%%%%%% conclusion %%%%%%%%%%%%%%%%%%%%%%%%%%%%%%%%%
%%%%%%%%%%%%%%%%%%%%%%%%%%%%%%%%%%%%%%%%%%%%%%%%%%%%%%%%%%%%%%%%%%%%%%%%%%%%%%%%
\section{Conclusion}
Parameter inference in stochastic kinetic models is an essential but challenging task that can, among other approaches, be tackled via gradient-based optimization. 
However, estimating gradients in the Gillespie SSA is not straightforward, owing to its discrete non-differentiable sampling steps. 

In this work, we adapted three gradient estimators from machine learning to gradient-based parameter inference in stochastic kinetic models: the GS-ST, the SF, and the AP~estimators.
We employed the estimators to evaluate gradients of steady-state observables (Fig.~\ref{fig:bima_overview_step}) and analyzed how the variance differs across the estimators (Fig.~\ref{fig:bima_var}). 
The SF and AP~estimators exhibit gradient variance that grows linearly with the length of the Gillespie trajectory. 
In contrast, the GS-ST estimator shows two distinct regimes, namely low asymptotic variance at high temperature and exponentially diverging variance at low temperature, with the latter emerging due to multiplicative accumulation of variance along the trajectory.
We further extended all three estimators to time-dependent observables by accounting for the gradient contribution of the waiting times (Fig.~\ref{fig:bima_overview_time}). 
This extension alters the variance scaling of the GS-ST estimator from constant variance to quadratic growth with time (at high temperature), partially offsetting its advantage of trajectory-length-independent variance (Fig.~\ref{fig:bima_var_time}).
Similarly, the AP estimator retains its linear variance scaling only for short times before crossing over to superlinear growth at long times.
In contrast, the SF estimator continues to exhibit linear variance scaling even for time-dependent observables, albeit with an increased slope.

Using the repressilator system as an example, we compared the performance of the estimators for parameter inference. 
For each estimator, we performed 50 inference tasks in which the ground-truth parameters were recovered from randomly initialized parameters by minimizing the loss function via stochastic gradient descent (Fig.~\ref{fig:repr_inf}).
We found that SF~recovered the ground-truth parameters most robustly, while GS-ST performed slightly worse and failed to converge for a small number of parameter sets.
Overall, AP~showed the weakest convergence. 
These performance differences were closely linked to the variance properties of the estimators (Fig.~\ref{fig:repr_var}). 
Analogously to the bimolecular association model, AP~exhibited higher variance than~SF across all parameter sets, consistent with its poorer convergence. 
GS-ST had lower variance than~SF for most parameter sets, but showed substantially larger variance in the non-converging optimizations. 
These cases corresponded to initializations at high binding affinity, a regime in which GS-ST performance suffered from diverging variance.

Our results indicate that, although GS-ST can provide low-variance gradient estimates in favorable parameter regimes, its performance degrades rapidly in more challenging ones. 
Increasing the GS-ST estimator temperature can restore convergence in such cases, but introduces additional bias and may impair inference elsewhere in parameter space. 
As a result, it is difficult to identify a single temperature~$\tau$ that yields robust performance across all parameter sets, a limitation that may restrict the applicability of GS-ST, particularly in high-dimensional models with complex bias–variance trade-offs. 
By contrast, the SF estimator appears to be more robust, as its performance is less sensitive to the specific parameter regime, although its applicability remains limited to relatively short Gillespie trajectories due to its linear variance scaling.

More broadly, our results identify large gradient variance as one of the central obstacles to robust gradient-based inference in stochastic simulators. 
Therefore, an important direction for future work is the adoption of additional variance-reduction strategies for GS-ST~\citep{paulus2021raoblackwellizing} and SF~\citep{greensmith2004variance}. 
A second goal might be to extend gradient estimation methods for the Gillespie SSA beyond fixed reaction networks to systems with dynamically changing network topologies, in which reactions can appear or disappear over time as new molecular species are formed. 
This is particularly important in complex biochemical systems, such as pools of reacting RNA oligomers, where the full network of chemical reactions cannot be enumerated in advance \citep{rosenberger2021selfassembly,burger2025vcg,harthkitzerow2026sequence}.
A third promising research direction arises from the SF~estimator: as it relies on evaluating the gradient of the log-likelihood, it naturally connects gradient-based inference to likelihood-based Bayesian inference. 
This raises the possibility of going from point estimation toward posterior inference, for example by using the score function in gradient-based Monte Carlo methods such as Hamiltonian Monte Carlo~\citep{duane1987hmc, neal2011hmc}.

Overall, our work shows that gradient estimators in the Gillespie SSA provide a viable route toward parameter inference in stochastic kinetic models, although the usefulness of a given estimator depends on its robustness across the model's parameter space. 
By systematically comparing different gradient estimators, we provide an overview of distinct approaches applied to differentiable stochastic simulation and highlight their respective advantages and limitations. 
In this way, our results help inform the choice of gradient estimators in future inference problems and provide a basis for the further development of robust gradient-based methods for stochastic kinetic models.

\begin{acknowledgments}
We thank Tom Magorsch, Nicole Hartman, Michael Kagan, and all members of the Gerland group for stimulating discussions. 
A.K. thanks the International Max Planck Research School for Intelligent Systems (IMPRS-IS) for support.
This work was supported by the Excellence Cluster ORIGINS which is funded by the German Research Foundation (DFG, Deutsche Forschungsgemeinschaft) under Germany’s Excellence Strategy -- EXC 2094-390783311.
\end{acknowledgments}

\section*{Data Availability}
The code used to generate and analyze the data presented in this study, including scripts to reproduce the figures, is available at \url{https://github.com/gerland-group/DifferentiableGillespie}. The implementation is primarily written in \texttt{Python} and uses \texttt{JAX}~\citep{jax2026github} for automatic differentiation.

\appendix
\section{Exact Solution of the Bimolecular Association Process}
\label{APsec:bima}

\paragraph*{Chemical Master Equation.} 
In bimolecular association, two molecular species A and B reversibly form a bimolecular complex A-B.
Let $N \in \{0, \dots, N^{\rm max}_{\rm A-B}\}$ denote the number of bimolecular complexes, where the maximum number of complexes is set by the total amount of molecules A and B, $N^{\rm max}_{\rm A-B} = \min(N_{\rm A}^{\rm tot}, N_{\rm B}^{\rm tot})$. 
The Chemical Master Equation (CME) for this system reads
\begin{align*}
    \frac{\partial p(n,t)}{\partial t}
    =\,& a_{\rm ass}(n-1)\,p(n-1,t) + a_{\rm dis}(n+1)\,p(n+1,t) \\
    & - a_{\rm ass}(n)\,p(n,t) - a_{\rm dis}(n)\,p(n,t) \,,
\end{align*}
where $p(n,t)$ denotes the probability for the system to contain $n$ bimolecular complexes at time $t$, and
\begin{align*}
    a_{\rm ass}(n) &= V^{-1}\bigl(N_A^{\mathrm{tot}}-n\bigr)\bigl(N_B^{\mathrm{tot}}-n\bigr) \, , \\
    a_{\rm dis}(n) &= k n \, ,
\end{align*}
denote the association and dissociation propensities, respectively.
As the dynamics are linear in the probability vector, \looseness=-1 $\bm{p}(t)\,=\,(p(0,t),\dots,p(N_{\max},t))$, we can express the CME equivalently as
\begin{align*}
    \frac{\partial \bm p}{\partial t} = \bm W \bm p \, ,
\end{align*}
with transition matrix elements
\begin{align*}
    W_{nm}
    =\,& V^{-1}\bigl(N_A^{\mathrm{tot}}-m\bigr)\bigl(N_B^{\mathrm{tot}}-m\bigr)\,(\delta_{m,n-1}-\delta_{m,n}) \\
    &+ km\,(\delta_{m,n+1}-\delta_{m,n}) \, .
\end{align*}

\paragraph*{Steady State Distribution and its Gradient.}
The steady-state distribution $\bm{p}_{\rm s.s.}$ satisfies
\begin{align*}
    \bm W \bm{p}_{\rm s.s.} = \bm 0 \, ,
\end{align*}
and can be obtained numerically as the right null vector of $\bm W$, normalized such that $\sum_n p_{\rm s.s.}(n)=1$.

Evaluating the gradient of the steady-state distribution via automatic differentiation is not possible: Finding the null vector of $\bm W$ requires a singular value decomposition (via \verb|jax.numpy.linalg.svd|) followed by extracting the eigenvector with eigenvalue zero, with the latter breaking automatic differentiation.

To determine the gradient of the steady-state probability distribution with respect to $k$, we derive an expression that relates the unknown gradient ${\rm d} \bm{p}_{\rm s.s.} / {\rm d} k$ to the known gradient of the transition matrix, ${\rm d} \bm{W} / {\rm d} k$. We find this relation by differentiating $\bm{W} \bm{p}_{\rm s.s.} = 0$ with respect to $k$, yielding
\begin{align*}
    \bm W\,\frac{{\rm d} \bm{p}_{\rm s.s.}}{{\rm d} k} = - \frac{{\rm d} \bm W}{{\rm d} k} \,\bm{p}_{\rm s.s.} \, .
\end{align*}
Since $\bm W$ is singular, we use the Moore--Penrose pseudoinverse $\bm W^+$ to construct a particular solution,
\begin{align*}
    \left( \frac{{\rm d} \bm{p}_{\rm s.s.}}{{\rm d} k} \right)_{\mathrm{part}}
    = - \bm W^+ \frac{{\rm d} \bm W}{{\rm d} k}\bm{p}_{\rm s.s.} \, .
\end{align*}
The general solution additionally allows a component in the nullspace of $\bm W$, i.e.\ a multiple of $\bm{p}_{\rm s.s.}$, to be added,
\begin{align*}
    \frac{{\rm d} \bm{p}_{\rm s.s.}}{{\rm d} k}
    = - \bm W^+ \frac{{\rm d} \bm W}{{\rm d} k} \bm{p}_{\rm s.s.} + \alpha \,\bm{p}_{\rm s.s.} \, .
\end{align*}
The normalization constraint $\sum_n p_n=1$ implies $\sum_n {\rm d} p_n / {\rm d} k =0$, which fixes $\alpha$,
\begin{align*}
    &0=\sum_n \frac{{\rm d} p_{\rm s.s.}(n)}{{\rm d} k}
    = -\sum_{n,i,j} W^+_{ni} \frac{{\rm d} W_{ij}}{{\rm d} k} p_{\rm s.s.}(j) + \alpha\sum_n p_{\rm s.s.}(n) \\
    &\Leftrightarrow \alpha=\sum_{n,i,j} W^+_{ni} \frac{{\rm d} W_{ij}}{{\rm d} k} p_{\rm s.s.}(j) \, .
\end{align*}
Given the gradient of the steady-state probability distribution, we can evaluate the gradient of any observable function of interest in steady state, for instance,
\begin{align*}
    \frac{{\rm d} \langle N \rangle}{{\rm d} k } = \frac{{\rm d} \sum_n n \, p_{\rm s.s.}(n)}{{\rm d} k } = \sum_n n \, \frac{{\rm d} p_{\rm s.s.}(n)}{{\rm d} k} \, .
\end{align*}

\paragraph*{Non-Stationary Distribution and its Gradient.}
To find the transient probability distribution (before the system has reached steady state), we solve the full time-dependent CME. The formal solution is obtained via the matrix exponential, 
\begin{align*}
    \bm{p}(t) = \exp \left[ \bm{W} t \right] \, \bm{p}(0) \, ,
\end{align*}
which can be computed numerically using the function \verb|jax.scipy.linalg.expm|. Since gradients of the matrix exponential can be evaluated via automatic differentiation in \verb|JAX|, the gradient ${\rm d}\bm{p} / {\rm d} k$ can be evaluated directly.

\section{Mean-Field Analysis of the Repressilator}
\label{APsec:repr_mf}

\begin{figure}
    \subfloat[]{\panel[width=\linewidth]{a}{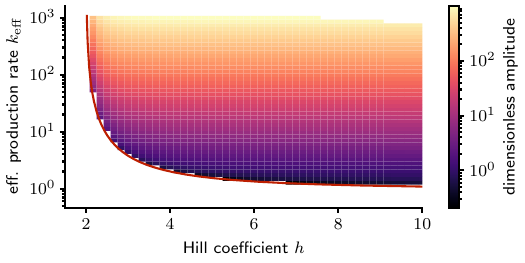}} \\[-5ex]
    \subfloat[]{\panel[width=\linewidth]{b}{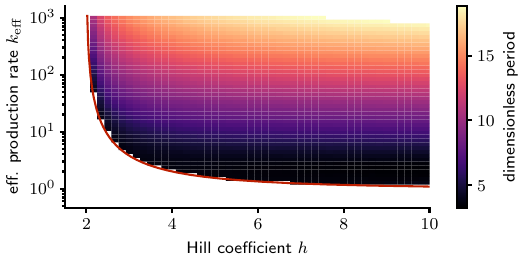}} \\[-5ex]
    \subfloat[]{\panel[width=0.49\linewidth]{c}{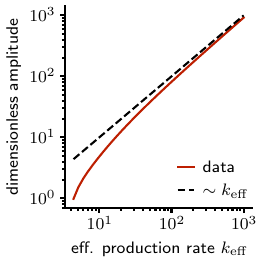}} \hfill
    \subfloat[]{\panel[width=0.49\linewidth]{d}{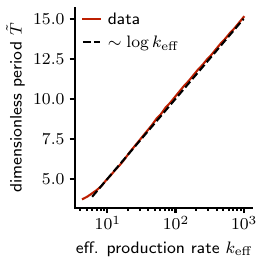}} \\[-3ex]
    \caption{
    Analysis of the chemical rate equations of the repressilator model.
    (a--b)~The mean-field repressilator model shows self-sustained oscillations with finite amplitude and period for Hill coefficient $h > 2$. The amplitude and period are largely insensitive to the choice of Hill coefficient, but depend on the effective production rate $k_{\rm eff}$.
    (c)~The dimensionless amplitude scales approximately linearly with $k_{\rm eff}$.
    (d)~The dimensionless period is proportional to $\ln k_{\rm eff}$.
    }
    \label{fig:repr_mf}
\end{figure}

The time evolution of the repressilator (considering only protein levels, and ignoring mRNA expression) is governed by the mean-field rate equations
\begin{align*}
    \frac{{\rm d} c_i}{{\rm d} t} = \frac{k_{\rm p}}{1 + \left( \frac{c_{i-1}}{K_{\rm d}} \right)^h} - k_{\rm d} c_i \, ,
\end{align*}
where $c_i$ denotes the concentration of protein species $i \in \{1,2,3\}$, $k_{\rm p}$ the protein production rate, $k_{\rm d}$ the degradation rate, $K_{\rm d}$ the dissociation constant, and $h$ the Hill coefficient~\citep{loinger2007stochastic}. 
Introducing the rescaled concentration $\tilde{c}_i=c_i/K_{\rm d}$ and rescaled time $\tilde{t} = t \cdot k_{\rm d}$, the governing equations take the dimensionless form
\begin{align*}
    \frac{{\rm d} \tilde{c}_i}{{\rm d} \tilde{t}} = \frac{k_{\rm eff}}{1 + \tilde{c}_{i-1}^h} - \tilde{c}_i \, ,
\end{align*}
with effective production rate $k_{\rm eff} = k_{\rm p} / ( K_{\rm d} \, k_{\rm d} )$.
The system shows self-sustained oscillations for $h > 2$ when $k_{\rm eff}$ exceeds a critical value at which the stationary solution becomes linearly unstable (Fig.~\ref{fig:repr_mf}a--b, see red line for the instability boundary \cite{bois2020repressilatornotes}). 
The amplitude and period of the oscillations are largely insensitive to the Hill coefficient for high $h$, but depend sensitively on $k_{\rm eff}$ (Fig.~\ref{fig:repr_mf}a--b). 

The amplitude scales approximately linearly with $k_{\rm eff}$ (Fig.~\ref{fig:repr_mf}c): 
the concentration of species $i$ increases until production and degradation balance, $k_{\rm eff} / (1 + \tilde{c}_{i-1}^h) = \tilde{c}_i$. 
Since the concentration of the suppressing species $i-1$ is low while species $i$ reaches its maximum, the balance simplifies to $\tilde{c}_i \approx k_{\rm eff}$, implying linear scaling of the amplitude with $k_{\rm eff}$. 
Reverting to dimensional quantities, this corresponds to $c \sim k_{\rm eff} K_{\rm d} = k_{\rm p}/k_{\rm d}$.
The period scales as $\tilde{T} \sim \ln k_{\rm eff}$ (Fig.~\ref{fig:repr_mf}d), since the concentration relaxes exponentially from its maximum value $\tilde{c}_{\rm max} \sim k_{\rm eff}$, corresponding to $T \sim k_{\rm d}^{-1} \ln (k_{\rm p} / (K_{\rm d} k_{\rm d}))$ in dimensional units.

These scaling relations determine which parameters can be inferred from trajectory data: if the absolute concentrations are known as a function of time (such that both the amplitude and period can be extracted), sufficient information is available to infer $k_{\rm p}/k_{\rm d}$ and $K_{\rm d}$, with $h$ assumed to be known.
Setting the degradation timescale as in the main text, $k_{\rm d} = 1$, this reduces to inferring $k_{\rm p}$ and $K_{\rm d}$.

\section{Bias-Driven Failure of Inference for GS-ST at High Temperature}
\label{APsec:repr_bias_gsst}
\begin{figure}
    \subfloat[]{\panel[width=0.49\linewidth]{a}{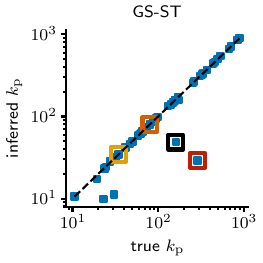}} \hfill
    \subfloat[]{\panel[width=0.49\linewidth]{b}{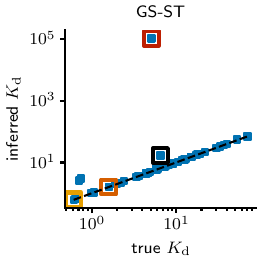}} \\[-7ex]
    \subfloat[]{\panel[width=0.49\linewidth]{c}{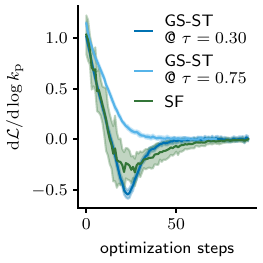}} \hfill
    \subfloat[]{\panel[width=0.49\linewidth]{d}{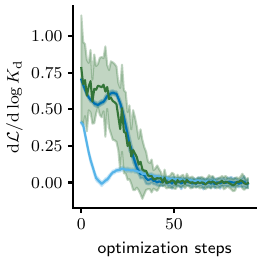}} \\[-7ex]
    \subfloat[]{\panel[width=0.49\linewidth]{e}{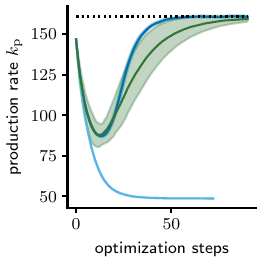}} \hfill
    \subfloat[]{\panel[width=0.49\linewidth]{f}{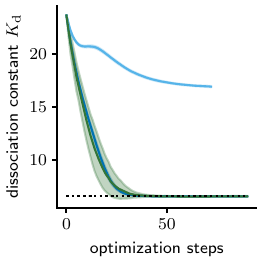}} \\[-3ex]
    \caption{
    Temperature-dependent performance of GS-ST for parameter inference in the repressilator model.
    (a--b)~Parameter inference via GS-ST at $\tau = 0.75$ succeeds for most parameter sets, including several that failed at $\tau = 0.30$ (highlighted by colored boxes), but a new subset of runs fails to converge.
    (c--d)~For one of the newly non-converging parameter sets (black boxes in panels a--b), the GS-ST gradient closely matches the reference gradient from the unbiased SF estimator at low temperature (dark blue), but deviates systematically at high temperature (light blue), indicating strong gradient bias.
    (e--f)~The gradient bias causes the optimization to converge to an incorrect optimum rather than the true parameters.
    In (c--f), lines and shaded regions show the mean and uncertainty across 20 independent optimization runs; in each run, gradients are estimated from 1000 stochastic trajectories at each optimization step.
    }
    \label{fig:repr_inf_hightau}
\end{figure}

As discussed in the main text, increasing the temperature of the GS-ST estimator reduces the variance of the gradient estimates, and is thus expected to restore convergence of parameter inference for the parameter sets that failed at $\tau = 0.3$.
Performing parameter inference for the same parameters as in Fig.~\ref{fig:repr_inf} using $\tau = 0.75$ confirms this expectation: inference succeeds for some of the previously challenging parameter sets (Fig.~\ref{fig:repr_inf_hightau}a--b, highlighted by colored boxes).
However, this comes at the cost of introducing a new set of non-converging inference runs, whose failure has a different origin: rather than large gradient variance, they arise from a systematic bias in the estimated gradient.

We illustrate this by following a typical optimization trajectory for one of the newly non-converging parameter sets. 
For low temperature, the GS-ST gradient closely resembles the true gradient obtained via the unbiased SF estimator (dark blue curve in Fig.~\ref{fig:repr_inf_hightau}c--d).
However, for high temperature, the GS-ST gradient deviates systematically from the true gradient 
(light blue curve in Fig.~\ref{fig:repr_inf_hightau}c--d). 
This bias is large enough to qualitatively alter the optimization landscape, causing the optimization to converge to an incorrect optimum rather than the true reference parameters (Fig.~\ref{fig:repr_inf_hightau}e--f).

These results illustrate a fundamental limitation of the GS-ST estimator: values of $\tau$ that ensure convergence in one region of parameter space may break inference in others, making it difficult to identify a single temperature that yields robust performance across the full parameter space.

\bibliography{mybib.bib}

@misc{anderson2017stochastic,
    author       = {Anderson, David F.},
    title        = {Lecture Notes on Stochastic Processes with Applications in Biology},
    year         = {2017},
    notes        = {p. 142},
    url = {https://u.math.biu.ac.il/~amirgi/SBA.pdf}
}

@article{arkin1998stochastic,
    author = {Adam Arkin and John Ross and Harley McAdams},
    title = {Stochastic Kinetic Analysis of Developmental Pathway Bifurcation in Phage $\lambda$-Infected {E}scherichia coli Cells},
    journal = {Genetics},
    volume = {149},
    number = {4},
    pages = {1633--1648},
    year = {1998},
    doi = {https://doi.org/10.1093/genetics/149.4.1633}
}

@inproceedings{arya2022automatic,
    title={Automatic Differentiation of Programs with Discrete Randomness},
    author={Gaurav Arya and Moritz Schauer and Frank Sch{\"a}fer and Christopher Vincent Rackauckas},
    booktitle={Conference on Neural Information Processing Systems},
    year={2022},
    url={https://openreview.net/forum?id=V22VeIZ9QU}
}

@inproceedings{arya2023diff,
    title={Differentiating Metropolis-Hastings to Optimize Intractable Densities},
    author={Gaurav Arya and Ruben Seyer and Frank Sch{\"a}fer and Kartik Chandra and Alexander K. Lew and Mathieu Huot and Vikash Mansinghka and Jonathan Ragan-Kelley and Christopher Vincent Rackauckas and Moritz Schauer},
    booktitle={ICML Workshop on Differentiable Almost Everything},
    year={2023},
    url={https://openreview.net/forum?id=2jag4Yatsz}
}

@misc{bois2020repressilatornotes,
  author = {Bois, Justin and Elowitz, Michael},
  title  = {The repressilator enables self-sustaining oscillations},
  year   = {2020},
  note   = {{L}ecture notes "Design Principles of Genetic Circuits", California Institute of Technology},
  url    = {https://be150.caltech.edu/2020/content/lessons/08_repressilator.html}
}

@article{brehmer:2019xox,
      author         = "Brehmer, Johann and Kling, Felix and Espejo, Irina and Cranmer, Kyle",
      title          = "{MadMiner: Machine learning-based inference for particle physics}",
      journal        = "Comput. Softw. Big Sci.",
      volume         = "4",
      year           = "2020",
      number         = "1",
      pages          = "3",
      doi            = "10.1007/s41781-020-0035-2",
}

@article{burger2025vcg,
    author = {Ludwig Burger and Ulrich Gerland},
    title = {Toward Stable Replication of Genomic Information in Pools of {RNA} Molecules},
    journal = {eLife},
    volume = {14},
    pages = {RP104043},
    year = {2025},
    doi = {https://doi.org/10.7554/eLife.104043}
}

@article{carter2024benchmark,
	title = {A benchmark {JWST} near-infrared spectrum for the exoplanet {WASP}-39 b},
    author = {Carter, Aarynn and others},
    journal = {Nat. Astron.},
	volume = {8},
    number = {8},
    pages = {1008--1019},
    year = {2024},
	doi = {https://doi.org/10.1038/s41550-024-02292-x},
}

@article{elowitz2000syntheticoscillatory,
    author = {Michael Elowitz and Stanislas Leibler},
    title = {A synthetic oscillatory network of transcriptional regulators},
    journal = {Nature},
    volume = {403},
    pages = {335-338},
    year = {2000},
    doi = {https://doi.org/10.1038/35002125},
}

@article{duane1987hmc,
    title = {Hybrid Monte Carlo},
    author = {Simon Duane and A.D. Kennedy and Brian J. Pendleton and Duncan Roweth},
    journal = {Phys. Lett. B},
    volume = {195},
    number = {2},
    pages = {216-222},
    year = {1987},
    doi = {https://doi.org/10.1016/0370-2693(87)91197-X},
    
}

@article{frank2022automatic,
    author = {Steven Frank},
    title = {Automatic Differentiation and the Optimization of Differential Equation Models in Biology},
    journal = {Front. Ecol. Evol.},
    volume = {10},
    pages = {1010278},
    year = {2022},
    doi = {https://doi.org/10.3389/fevo.2022.1010278}
}

@article{froehlich2016inference,
    author = {Fabian Fröhlich and Philipp Thomas and Atefeh Kazeroonian and Fabian Theis and Ramon Grima and Jan Hasenauer},
    title = {Inference for Stochastic Chemical Kinetics Using Moment Equations and System Size Expansion},
    journal = {PLoS Comp. Biol.},
    volume = {12},
    number = {7},
    pages = {e1005030},
    year = {2016},
    doi = {https://doi.org/10.1371/journal.pcbi.1005030},
}

@article{froehlich2017scalable,
    author = {Fröhlich, Fabian and Kaltenbacher, Barbara and Theis, Fabian and Hasenauer, Jan},
    title = {Scalable Parameter Estimation for Genome-Scale Biochemical Reaction Networks},
    journal = {PLoS Comp. Biol.},
    volume = {13},
    number = {1},
    pages = {e1005331},
    year = {2017},
    doi = {https://doi.org/10.1371/journal.pcbi.1005331}
}

@article{gavrikov2026neutrinosbi,
	title = {Simulation-based inference for precision neutrino physics through neural {Monte} {Carlo} tuning},
	author = {Gavrikov, Arsenii and others and {JUNO-Italia Consortium}},
    journal = {Commun. Phys.},
    volume = {9},
	number = {1},
    pages = {63},
    doi = {https://doi.org/10.1038/s42005-026-02499-6},
	year = {2026},
}

@article{gillespie1977exact,
    author = {Daniel Gillespie},
    title = {Exact Stochastic Simulation of Coupled Chemical Reactions},
    journal = {J. Phys. Chem.},
    volume = {81},
    number = {25},
    pages = {2340--2361},
    year = {1977},
    doi = {https://doi.org/10.1021/j100540a008},
}

@article{golightly2011bayesian,
    author = {Andrew Golightly and Darren Wilkinson},
    title = {Bayesian Parameter Inference for Stochastic Biochemical Network Models using Particle Markov Chain Monte Carlo},
    journal = {Interface Focus},
    volume = {1},
    number = {6},
    pages = {807--820},
    year = {2011},
    doi = {https://doi.org/10.1098/rsfs.2011.0047},
}

@article{greensmith2004variance,
    author = {Greensmith, Evan and Bartlett, Peter L. and Baxter, Jonathan},
    title = {Variance Reduction Techniques for Gradient Estimates in Reinforcement Learning},
    journal = {J. Mach. Learn. Res.},
    volume = {5},
    pages = {1471–1530},
    year = {2004},
    url = {https://www.jmlr.org/papers/volume5/greensmith04a/greensmith04a.pdf}
}

@book{gumbel1954statistical,
    author    = {Emil J. Gumbel},
    title     = {Statistical Theory of Extreme Values and Some Practical Applications: A Series of Lectures},
    year      = {1954},
    publisher = {U.S. Government Printing Office},
    url = {https://ntrl.ntis.gov/NTRL/dashboard/searchResults/titleDetail/PB175818.xhtml},
}

@book{harris2012digital,
    author = {David Harris and Sarah Harris},
    title = {Digital Design and Computer Architecture},
    year = {2012},
    publisher = {Elsevier},
    edition = {2},
}

@article{harthkitzerow2026sequence,
    author = {Johannes Harth-Kitzerow and Tobias Göppel and Ludwig Burger and Torsten Enßlin and Ulrich Gerland},
    title = {Sequence motif dynamics in {RNA} pools},
    journal = {Phys. Rev. E},
    volume = {113},
    pages = {024407},
    year = {2026},
    doi = {https://doi.org/10.1103/y5qv-qm7n},
}

@misc{heinrich2026differentiablequantum,
      title={Differentiable quantum-trajectory simulation of {L}indblad dynamics for {QGP} transport-coefficient inference}, 
      author={Lukas Heinrich and Tom Magorsch},
      eprint = {2601.14399},
      archivePrefix = {arXiv},
      year={2026},
      url={https://arxiv.org/abs/2601.14399}, 
}

@inproceedings{jang2017categorical,
    title={Categorical Reparameterization with Gumbel-Softmax},
    author={Eric Jang and Shixiang Gu and Ben Poole},
    booktitle={International Conference on Learning Representations},
    year={2017},
    url={https://openreview.net/forum?id=rkE3y85ee}
}

@software{jax2026github,
    author = {James Bradbury and Roy Frostig and Peter Hawkins and Matthew James Johnson and Yash Katariya and Chris Leary and Dougal Maclaurin and George Necula and Adam Paszke and Jake Vander{P}las and Skye Wanderman-{M}ilne and Qiao Zhang},
    title = {{JAX}: composable transformations of {P}ython+{N}um{P}y programs},
    url = {http://github.com/jax-ml/jax},
    version = {0.7.1},
    year = {2026},
}

@misc{kagan2023branches,
    author = {Kagan, Michael and Heinrich, Lukas},
    title = {Branches of a Tree: Taking Derivatives of Programs with Discrete and Branching Randomness in High Energy Physics},
    eprint = {2308.16680},
    archivePrefix = {arXiv},
    year = {2023},
    url = {https://arxiv.org/abs/2308.16680},
}

@article{khan2021bayesian,
    author = {Aminul Islam Khan and Md Muhtasim Billah and Chunhua Ying and Jin Liu and Prashanta Dutta},
    title = {Bayesian Method for Parameter Estimation in Transient Heat Transfer Problem},
    journal = {Int. J. Heat Mass Transf.},
    volume = {166},
    pages = {120746},
    year = {2021},
    doi = {https://doi.org/10.1016/j.ijheatmasstransfer.2020.120746},
}

@inproceedings{kingma2014autoencoding,
    author = {Diederik Kingma and Max Welling},
    title = {Auto-Encoding Variational Bayes},
    booktitle = {International Conference on Learning Representations},
    year = {2014},
    url={https://openreview.net/forum?id=33X9fd2-9FyZd},
}

@misc{kofler2025flexible,
      title={Flexible Gravitational-Wave Parameter Estimation with Transformers}, 
      author={Annalena Kofler and Maximilian Dax and Stephen R. Green and Jonas Wildberger and Nihar Gupte and Jakob H. Macke and Jonathan Gair and Alessandra Buonanno and Bernhard Schölkopf},
      year={2025},
      eprint={2512.02968},
      archivePrefix={arXiv},
      url={https://arxiv.org/abs/2512.02968}, 
}

@article{komorowski2009bayesian,
    author = {Michał Komorowski and Bärbel Finkenstädt and Claire Harper and David Rand},
    title = {Bayesian Inference of Biochemical Kinetic Parameters using the Linear Noise Approximation},
    journal = {BMC Bioinformatics},
    volume = {10},
    pages = {343},
    year = {2009},
    doi = {https://doi.org/10.1186/1471-2105-10-343},
}

@article{komorowski2011sensitivity,
    author = {Michał Komorowski and Maria Costa and David Rand and Michael Stumpf},
    title = {Sensitivity, robustness, and identifiability in stochastic chemical kinetics models},
    journal = {Proc. Natl. Acad. Sci. USA},
    volume = {108},
    pages = {8645 -- 8650},
    year = {2011},
    doi = {https://doi.org/10.1073/pnas.1015814108}
}

@article{lueck2016generalized,
    author = {Alexander Lück and Verena Wolf},
    title = {Generalized Method of Moments for Estimating Parameters of Stochastic Reaction Networks},
    journal = {BMC Syst. Biol.},
    volume = {10},
    pages = {98},
    year = {2016},
    doi = {https://doi.org/10.1186/s12918-016-0342-8}
}

@article{loinger2007stochastic,
    author = {Adiel Loinger and Ofer Biham},
    title = {Stochastic simulations of the repressilator circuit},
    journal = {Phys. Rev. E},
    volume = {76},
    pages = {051917},
    year = {2007},
    doi = {https://doi.org/10.1103/PhysRevE.76.051917}
}

@inproceedings{maddison2017discrete,
  author = {Chris J. Maddison and Andriy Mnih and Yee Whye Teh},
  title = {The Concrete Distribution: {A} Continuous Relaxation of Discrete Random Variables},
  booktitle = {International Conference on Learning Representations},
  year = {2017},
  url = {https://openreview.net/forum?id=S1jE5L5gl},
}

@article{mitra2020parameter,
    author = {Eshan Mitra and William Hlavacek},
    title = {Parameter Estimation and Uncertainty Quantification for Systems Biology Models},
    journal = {Curr. Opin. Syst. Biol.},
    volume = {18},
    pages = {9--18},
    year = {2020},
    doi = {https://doi.org/10.1016/j.coisb.2019.10.006},
}

@article{mohamed2020montecarlo,
  author  = {Shakir Mohamed and Mihaela Rosca and Michael Figurnov and Andriy Mnih},
  title   = {Monte Carlo Gradient Estimation in Machine Learning},
  journal = {J. Mach. Learn. Res.},
  volume  = {21},
  number  = {132},
  pages   = {1--62},
  year    = {2020},
  url     = {http://jmlr.org/papers/v21/19-346.html}
}

@misc{mottes2026gradient,
	author = {Mottes, Francesco and Zhu, Qian-Ze and Brenner, Michael},
    title = {Gradient-based optimization of exact stochastic kinetic models},
    eprint = {2601.14183},
    archivePrefix = {arXiv},
	year = {2026},
    url = {http://arxiv.org/abs/2601.14183},
}

@incollection{neal2011hmc,
    author = {Radford Neal},
    title = {{MCMC} Using Hamiltonian Dynamics},
    booktitle = {Handbook of Markov Chain Monte Carlo},
    publisher = {Chapman and Hall/CRC},
    year = 2011,
    doi = {https://doi.org/10.1201/b10905},      
}

@inproceedings{paulus2021raoblackwellizing,
    title={{R}ao-{B}lackwellizing the {S}traight-{T}hrough {G}umbel-{S}oftmax Gradient Estimator},
    author={Max B Paulus and Chris J. Maddison and Andreas Krause},
    booktitle={International Conference on Learning Representations},
    year={2021},
    url={https://openreview.net/forum?id=Mk6PZtgAgfq}
}

@article{potvin2016synchronous,
    author = {Laurent Potvin-Trottier and Nathan Lord and Glenn Vinnicombe and Johan Paulsson},
    title = {Synchronous long-term oscillations in a synthetic gene circuit},
    journal = {Nature},
    volume = {538},
    pages = {514--517},
    year = {2016},
    doi = {https://doi.org/10.1038/nature19841}
}

@article{rao2016nonequilibrium,
    author = {Riccardo Rao and Massimiliano Esposito},
    title = {Nonequilibrium Thermodynamics of Chemical Reaction Networks: Wisdom from Stochastic Thermodynamics},
    journal = {Phys. Rev. X},
    volume = {6},
    pages = {041064},
    year = {2016},
    doi = {https://doi.org/10.1103/PhysRevX.6.041064}
}

@inproceedings{rezende2014stochastic,
    author = {Rezende, Danilo Jimenez and Mohamed, Shakir and Wierstra, Daan},
    title = {Stochastic backpropagation and approximate inference in deep generative models},
    year = {2014},
    booktitle = {International Conference on Machine Learning},
    url = {https://dl.acm.org/doi/10.5555/3044805.3045035}
}

@article{rijal2025diffablegillespie,
    title = {A differentiable Gillespie algorithm for simulating chemical kinetics, parameter estimation, and designing synthetic biological circuits},
    author = {Rijal, Krishna and Mehta, Pankaj},
    journal = {eLife},
    volume = {14},
    pages = {RP103877},
    year = {2025},
    doi = {https://doi.org/10.7554/eLife.103877},
}

@article{rosenberger2021selfassembly,
    author = {Joachim Rosenberger and Tobias Göppel and Patrick Kudella and Dieter Braund and Ulrich Gerland and Bernhard Altaner},
    title = {Self-Assembly of Informational Polymers by Templated Ligation},
    journal = {Phys. Rev. X},
    volume = {11},
    pages = {031055},
    year = {2021},
    doi = {https://doi.org/10.1103/PhysRevX.11.031055},
}

@article{rubner2000earthmoversdistance,
    author = {Yossi Rubner and Carlo Tomasi and Leonidas Guibas},
    title = {The {E}arth {M}over's {D}istance as a metric for image retrieval},
    journal = {Int. J. Comput. Vis.},
    volume = {40},
    pages = {99--121},
    year = {2000},
    doi = {https://doi.org/10.1023/A:1026543900054}
}

@article{ruess2017sensitivity,
    author = {Jakob Ruess and Heinz Koeppl and Christoph Zechner},
    title = {Sensitivity estimation for stochastic models of biochemical reaction networks in the presence of extrinsic variability},
    journal = {J. Chem. Phys.},
    volume = {146},
    pages = {124122},
    year = {2017},
    doi = {https://doi.org/10.1063/1.4978940}
}

@article{schnoerr2017approximationandinference,
    author = {David Schnoerr and Guido Sanguinetti and Ramon Grima},
    title = {Approximation and inference methods for stochastic biochemical kinetics -- a tutorial review},
    journal = {J. Phys. A: Math. Theor.},
    volume = {50},
    pages = {093001},
    year = {2017},
    doi = {https://doi.org/10.1088/1751-8121/aa54d9}
}

@article{schnoerr2015comparison,
    author = {David Schnoerr and Guido Sanguinetti and Ramon Grima},
    title = {Comparison of different moment-closure approximations for stochastic chemical kinetics},
    journal = {J. Chem. Phys.},
    volume = {143},
    pages = {185101},
    year = {2015},
    doi = {https://doi.org/10.1063/1.4934990}
}

@misc{schulman2016gradient,
  title={Gradient Estimation Using Stochastic Computation Graphs}, 
  author={John Schulman and Nicolas Heess and Theophane Weber and Pieter Abbeel},
  eprint={1506.05254},
  archivePrefix={arXiv},
  year={2016},
  url={https://arxiv.org/abs/1506.05254}, 
}

@article{stapor2018optimization,
    author = {Paul Stapor and Fabian Fröhlich and Jan Hasenauer},
    title = {Optimization and Profile Calculation of {ODE} Models using Second Order Adjoint Sensitivity Analysis},
    journal = {Bioinformatics},
    volume = {34},
    number = {13},
    pages = {i151 -- i159},
    year = {2018},
    doi = {https://doi.org/10.1093/bioinformatics/bty230},
}

@article{toni2009abc,
    author = {Tina Toni and David Welch and Natalja Strelkowa and Andreas Ipsen and Michael Stumpf},
    title = {Approximate Bayesian computation scheme for parameter inference and model selection in dynamical systems},
    journal = {J. R. Soc. Interface},
    volume = {6},
    pages = {187--202},
    year = {2009},
    doi = {https://doi.org/10.1098/rsif.2008.0172},
}

@misc{vilar2026exact,
      title={Exact Discrete Stochastic Simulation with Deep-Learning-Scale Gradient Optimization}, 
      author={Jose M. G. Vilar and Leonor Saiz},
      eprint={2602.19775},
      archivePrefix={arXiv},
      year={2026},
      url={https://arxiv.org/abs/2602.19775}, 
}

@article{warne2019simulation,
    author = {David Warne and Ruth Baker and Matthew Simpson},
    title = {Simulation and Inference Algorithms for Stochastic Biochemical Reaction Networks: From Basic Concepts to State-of-the-Art},
    journal = {J. R. Soc. Interface.},
    volume = {16},
    number = {151},
    pages = {20180943},
    year = {2019},
    doi = {https://doi.org/10.1098/rsif.2018.0943},
}

@article{williams1992simple,
	title = {Simple statistical gradient-following algorithms for connectionist reinforcement learning},
	author = {Williams, Ronald},
    journal = {Mach. Learn.},
    volume = {8},
    number = {3},
    pages = {229--256},
    year = {1992},
	doi = {https://doi.org/10.1007/BF00992696},
}

@book{zheng2018feature,
  title={Feature Engineering for Machine Learning: Principles and Techniques for Data Scientists},
  author={Zheng, Alice and Casari, Amanda},
  year={2018},
  publisher={O'Reilly Media, Inc.},
  address={Sebastopol, CA},
  edition={1},
  isbn={978-1-4919-5324-2}
}

@article{zechner2012momentbased,
    author = {Christoph Zechner and Jakob Ruess and Peter Krenn and Serge Pelet and Matthias Peter and John Lygeros and Heinz Koeppl},
    title = {Moment-based inference predicts bimodality in transient gene expression},
    journal = {Proc. Natl. Acad. Sci. USA},
    volume = {109},
    pages = {8340--8345},
    year = {2012},
    doi = {https://doi.org/10.1073/pnas.1200161109}
}

\end{document}